\documentclass[a4paper,number,compress]{elsarticle}

\usepackage[australian]{babel}
\usepackage[T1]{fontenc} 
\usepackage{lmodern}
\usepackage{amssymb}
\usepackage{amsmath}
\usepackage{mathrsfs} 
\usepackage{textcomp} 

\usepackage[breaklinks,plainpages=false,pdfpagelabels]{hyperref}
\hypersetup{colorlinks,citecolor=blue,filecolor=blue,linkcolor=blue,urlcolor=blue}

\usepackage{graphicx}
\usepackage[caption=false]{subfig}
\usepackage{nicefrac}
\usepackage{fixmath}
\usepackage{fixltx2e} 
\usepackage{array}
\usepackage[printonlyused,withpage]{acronym} 
\usepackage{microtype} 

\newcommand{\mat}[1]{\mathbold{#1}} 
\newcommand{\matGreek}[1]{\mat{#1}} 
\newcommand{\itGreek}[1]{\mathit{#1}} 

\makeatletter
\def\imod#1{\allowbreak\mkern10mu({\operator@font mod}\,#1)}
\makeatother

\newcommand{\placement}{htbp}

\newcommand{\primeSymbol}{\textit{p}}

\newcommand{\firstIndex}{j}

\newcommand{\eqnTag}{Eq.}
\newcommand{\eqnsTag}{Eqs}
\newcommand{\EqnTag}{\eqnTag}
\newcommand{\figTag}{Fig.}

\newcommand{\xSymbol}{x}
\newcommand{\ySymbol}{y}
\newcommand{\setSeparator}{:}

\newcommand{\numberSymbol}{N}

\newcommand{\transformDiscreteRadon}{\mathcal{R}}
\newcommand{\transformMojette}{\mathcal{M}}

\newcommand{\wholeNumbers}{\mathbb{N}_0}

\newcommand{\realNumbers}{\mathbb{R}}
\newcommand{\integerNumbers}{\mathbb{Z}}

\journal{Signal Processing}

\begin{document}
 
\begin{frontmatter}
 \title{Fast Mojette Transform for Discrete Tomography}

 \author[mon]{Shekhar Chandra}
 \ead{Shekhar.Chandra@monash.edu}
 \author[mon,poly]{Nicolas Normand}
 \ead{Nicolas.Normand@monash.edu or nicolas.normand@polytech.univ-nantes.fr}
 \author[anu]{Andrew Kingston}
 \ead{Andrew.Kingston@anu.edu.au}
 \author[poly]{Jeanpierre Gu\'edon}
 \ead{jeanpierre.guedon@polytech.univ-nantes.fr}
 \author[mon]{Imants Svalbe}
 \ead{Imants.Svalbe@monash.edu}

 \address[mon]{School of Physics, Monash University, Australia.}
 \address[poly]{IRCCyN-IVC, \'{E}cole polytechnique de l'Universit\'{e} de Nantes, France.}
 \address[anu]{Department of Applied Mathematics, Australian National University.}

 \begin{abstract}
 A new algorithm for reconstructing a \acl*{2D} object from a set of \acl*{1D} projected views is presented that is both computationally exact and experimentally practical. The algorithm has a computational complexity of $O(n\log_2 n)$ with $n = N^2$ for an $N\times N$ image, is robust in the presence of noise and produces no artefacts in the reconstruction process, as is the case with conventional tomographic methods. The reconstruction process is approximation free because the object is assumed to be discrete and utilises fully discrete Radon transforms. Noise in the projection data can be suppressed further by introducing redundancy in the reconstruction. The number of projections required for exact reconstruction and the response to noise can be controlled without comprising the digital nature of the algorithm. The digital projections are those of the \acl*{MT}, a form of discrete linogram. A simple analytical mapping is developed that compacts these projections exactly into symmetric periodic slices within the \acl*{DFT}. A new digital angle set is constructed that allows the periodic slices to completely tile all of the objects Discrete Fourier space. Techniques are proposed to acquire these digital projections experimentally to enable fast and robust \acl*{2D} reconstructions.
 \end{abstract}

\end{frontmatter}

\section{Introduction}
A common problem in science is to determine the internal structure of an object from indirect measurements. Tomography is concerned with the recovery or reconstruction of the internal structure of an object from its projected ``views'' or projections. In practical terms, projections provide density profiles of an object $f$ at different angles $\theta$, either from absorption or transmission of, for example, x-rays. The profiles are directly proportional to the attenuation of the x-ray intensity within the object. This attenuation is proportional to the magnitude of the line integral 
\begin{equation}
 \mu_\theta(t) = \int_{L_{\theta,t}} f(\xSymbol,\ySymbol)\: d\ell, \label{eqn::Projection}
\end{equation}
at each translate $t$ and where $x, y$ are \ac{2D} spatial coordinates. The curves $L_{\theta,t}$ (usually) form a set of parallel lines $t\in \realNumbers$ (with compact support $D$) within the projection $\mu_\theta(t)$ having the measure $d\ell$ on $\mathbb{R}^2$~\citep{RammKatsevich}.

An exact reconstruction of the object can then be made when the set of projections $\matGreek{\mu}$ is given as
\begin{equation}
 \matGreek{\mu} = \left\{\mu_\theta \setSeparator \theta \in [0,\pi)\right\}, \label{eqn::RT}
\end{equation}
where the functional notation for $\mu_\theta(t)$ has been dropped for convenience and it is assumed that $\mu_\theta=\mu_{\theta+\pi}$ for simplicity\footnote{A discussion of the effects of scattering and beam hardening in polychromatic x-ray beams is beyond the scope of this paper.}. Equations~\eqref{eqn::Projection} and~\eqref{eqn::RT} define the \ac{2D} \ac{RT} of the object first developed for ray lines by Radon~\citep{Radon} and Funk~\citep{Funk1915} when using ``great circles''. Note that the set~\eqref{eqn::RT} is uncountably infinite, therefore requiring an infinite number of projections for an exact or unambiguous reconstruction of the continuous function $f$~\citep{Smith1977}. Thus, the reconstruction is always ill-posed when using the \ac{RT}, since only a finite number of measurements can be made practically~\citep{Logan1975}. 

The conventional solution is to acquire a large number of projections (usually) at evenly spaced angles. This has practical implications on the quality of the reconstructed image~\citep{Katz}. The uniqueness of the reconstruction directly relates to the artefacts known as Ghosts or ``invisible distributions''~\citep{Bracewell1954}. Ghosts are always present in reconstructions when using the methods based on this classical \ac{RT}~\citep{Louis1981}.

Discrete \acp{RT} attempt to resolve this issue by assuming that the object is partitioned into discrete elements, so that an exact reconstruction may be computed in the discrete domain with a finite number of projections. The inverse problem of \ac{RT} is then no longer ill-posed and therefore free from reconstruction artefacts. Notable discrete \acp{RT} include the arbitrary curve block circulant discrete \ac{RT}~\citep{Beylkin, Kelley1992}, the $d$-lines discrete \ac{RT}~\citep{Gotz1996, Brady1998} and the Fast Slant Stack~\citep{Averbuch2001}. A short review of these methods, which are not required for this work, is provided in~\ref{sec::DRTs}. However, most discrete methods tend to suffer from either experimental limitations or high computational complexity. This paper presents a discrete approach that is both experimentally and computationally viable that utilises the \ac{MT}, a discrete \ac{RT} first constructed by Gu\'edon \emph{et al.}~\citep{Guedon1995}.

\subsection{Mojette Transform}
\begin{figure}[\placement]
 \centering
 \includegraphics[width=0.95\textwidth]{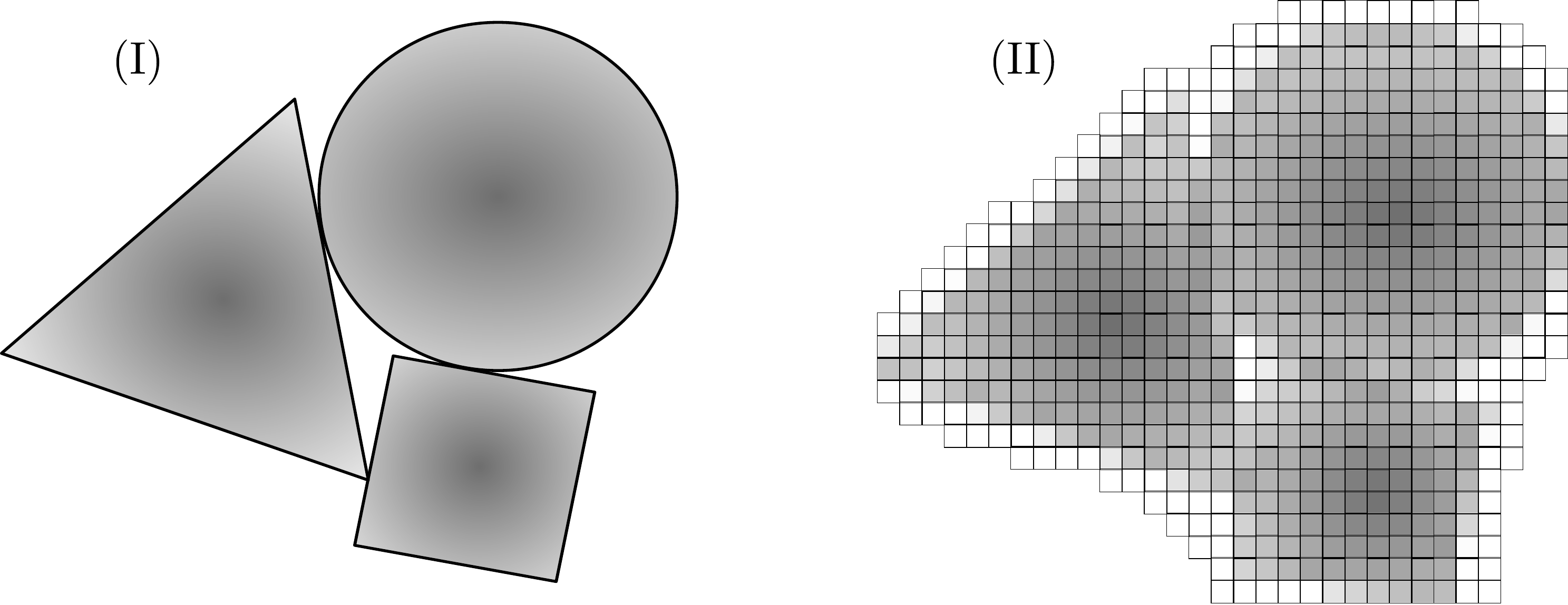}
 \caption{An example of a continuous object (I) and its digital representation (II). The digital version is deliberately represented in low resolution for illustrative purposes. Note that the object and its underlying array need not be rectangular.}
 \label{fig::Digital}
\end{figure}

Let the continuous object $f$ be approximated digitally within a point-lattice or digital array $\Lambda$ (see \figTag~\ref{fig::Digital}). 
This is a reasonable approximation since the reconstruction will eventually be represented and processed digitally. Also, let the rays form part of a parallel beam, in order to simplify what follows. Then, analogous to \eqnsTag~(\ref{eqn::Projection}) and~(\ref{eqn::RT}), the projections within the \ac{MT} are defined as
\begin{equation}
 \mu_{\theta_{pq}}(t) = \sum_{\Gamma_{t,\theta_{pq}}} f(\xSymbol,\ySymbol), \label{eqn::MTProjection}
\end{equation}
where $\theta_{pq} = \tan^{-1}(\nicefrac{q}{p})$ (or the vectors $[q,p]$), with respect to the array $\Lambda$ so that $p,q\in \integerNumbers$ and the $\gcd(p,q)=1$ (i.e. $p$ and $q$ are coprime). The fractions $\nicefrac{q}{p}$ are known as Farey fractions and are also related to the distribution of prime numbers~\citep{Franel1924, Landau1924}. These rational angle projections have a form equivalent to that of the discrete linogram, i.e. the number of bins $B$ is dependent on the angle $\theta_{pq}$ as
\begin{eqnarray}\label{eqn::MT_BinNumber}
 B = |p|(Q-1) + |q|(P-1) + 1,
\end{eqnarray}
for each projection of a rectangular $P\times Q$ image. This has implications on the choice of detector resolution and geometry, which are also discussed later in this work.

\begin{figure}[\placement]
 \centering
 \includegraphics[width=0.95\textwidth]{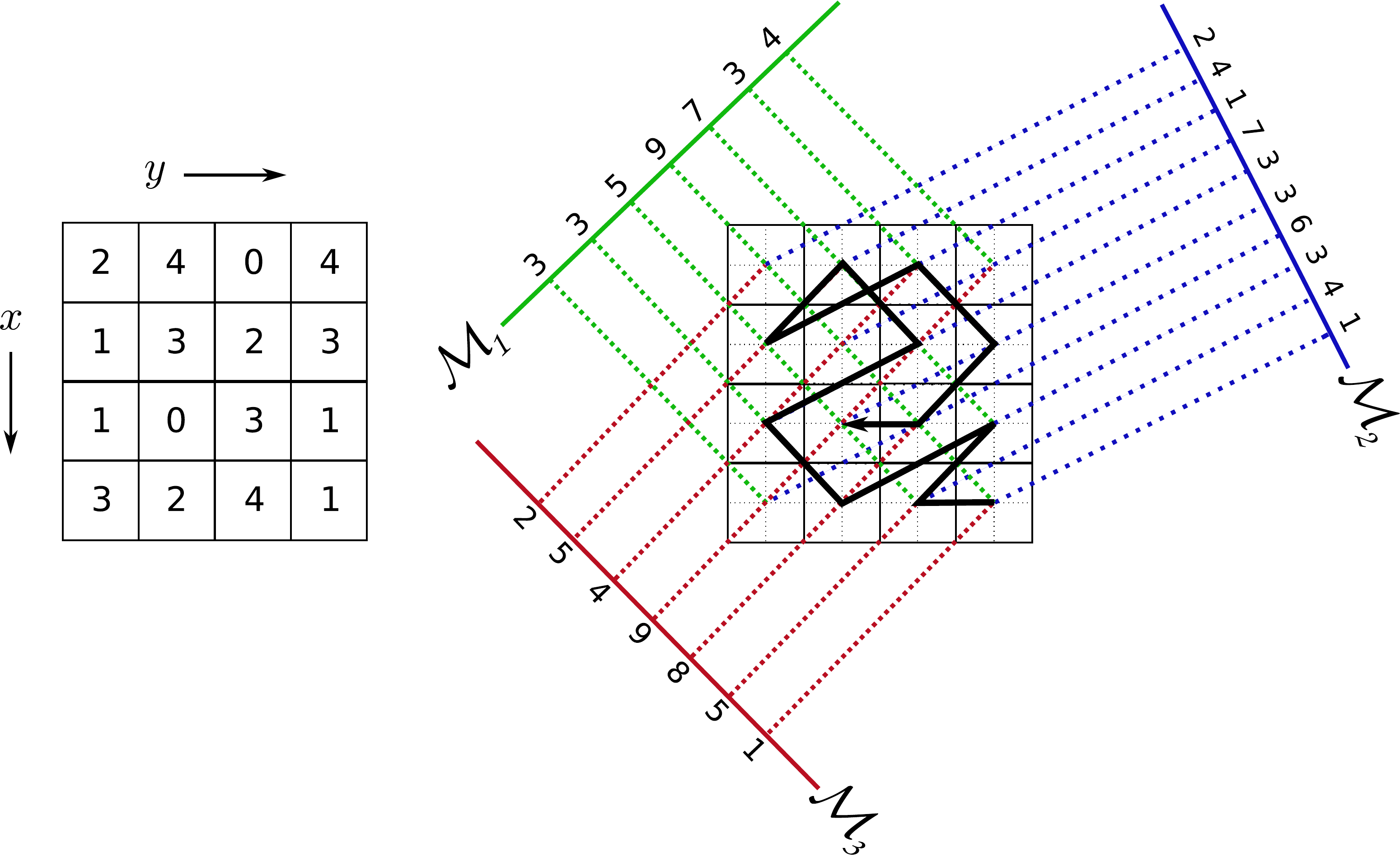}
 \caption{An example of a \acl*{MT} for a discrete image of size $4\times 4$ using the three projections $[1,1]$, $[1,-1]$ and~$[1,-2]$. The bold lines within the right-hand grid shows a possible reconstruction path using a corner-based inversion method~\citep{Normand2006}.}
\label{fig::MT}
\end{figure}

The definition of the lines $\itGreek{\Gamma}_{t,\theta_{pq}}$ fall into one of two classes, depending on the desired sampling being fully discrete or as intensities distributed over the area between lattice points. In the \ac{DPM}, the pixel is summed to its corresponding bin if and only if the line passes through the centre of the pixel. The lines $\Gamma_{t,\theta_{pq}}$ then form a set of non-periodic and parallel discrete lines
\begin{equation}\label{eqn::DiracPM}
\Gamma_{t,\theta_{pq}} = 
 \begin{cases}
 t = qy-px & \text{if}\quad \nicefrac{q}{p} \geqslant 0\\ 
 t = px-qy & \text{if}\quad \nicefrac{q}{p} < 0\\
 \end{cases},
\end{equation}
with $t\in \integerNumbers$ of an object having convex support. Convex support simply means that space does not have any local singularities. In interpolated or fractional \acp{MT}, the pixels are summed according to the length of the pixel intersected by the lines $\Gamma_{t,\theta_{pq}}$ and can be used to approximate continuous objects~\citep{Guedon1997}. Arbitrarily complicated pixel models are also possible using splines of high order. The \ac{DPM} is assumed for the duration of this paper. \figTag~\ref{fig::MT} shows a simple example of a \ac{MT} for a $4\times 4$ image using three projections.

The \ac{MT} may be inverted or reconstructed exactly according to one of two, albeit practically unsatisfactory, definitions of the Mojette projection set. Firstly, one acquires the projections at all possible angles $\theta_{pq}$ for a given image size, i.e. $\theta_{pq}$ is the set of all rational vectors of order $\numberSymbol$ in all octants of the half-plane for an $\numberSymbol\times\numberSymbol$ image. This definition of the \ac{MT} can be inverted exactly by smearing (or back-projecting) all the projections within this Mojette set~\citep{Servieres2005a}. Although the number of projections are finite in number, in contrast to the \ac{RT}, it still utilises a large number projections since $\theta_{pq}$ is large for modest $\numberSymbol$. For example, the inversion of a $128\times 128$ image requires a total of 20088 projections and is not computationally viable.

An alternative and more flexible definition of the Mojette projection set was proposed by Katz~\citep{Katz}. Since one needs only as many equations (or translates in this case) as pixels in order to reconstruct exactly, Katz pointed out that a set satisfying
\begin{eqnarray}
 P \leqslant \sum_{\firstIndex=0}^{M-1} |p_\firstIndex| \quad\ \text{or}\quad\  Q \leqslant \sum_{\firstIndex=0}^{M-1} |q_\firstIndex|, \label{eqn::MT_KatzCriterion}
\end{eqnarray}
where $M$ is the number of projections of a $P\times Q$ image, is theoretically sufficient to obtain an exact reconstruction. Normand \emph{et al.}~\citep{NormandEtAl} extended the concept to discrete objects within arbitrary convex regions. Although this new definition substantially reduces the number of projections required for an exact reconstruction, it requires new and more sophisticated inversion methods, as the original back-projection approach is no longer applicable. A number of schemes have been proposed, including a Conjugate Gradient method~\citep{Servieres2005} and a Geometric Graph approach~\citep{Normand2006}. The former is robust in the presence of noise but is not suitably convergent, i.e. an appropriate pre-conditioner is yet to be found, and the latter is very sensitive to noise.

The \ac{MT} was initially constructed to aid work that modelled the human vision system~\citep{Guedon1995}. Since then, applications of the \ac{MT} include image encoding and network transmission~\citep{NormandEtAl, Guedon1997}, as well as work on Asynchronous Transfer Mode (ATM), data integrity~\citep{Philippe1997a, Terrien1997}, packet networks via an $n$-dimensional \ac{MT}~\citep{Parrein2001}, lossless networking~\citep{Verbert2002a} and in scalable multimedia distribution~\citep{Guedon2001}. Image watermarking using a Fourier-based method and the \ac{MT} has also been developed~\citep{Autrusseau2002a,Autrusseau2002} as well as another scheme for crypto-marking and transmission over the Internet of medical images~\citep{AutrusseauEtAl, Babel2005}. Finally, the \ac{MT} has been applied to \ac{CT}~\citep{Servieres2003, Servieres2005a, Chandra2008, Fayad2008}. See Gu\'edon \emph{et al.}~\citep{Guedon2009} for a thorough review of the \ac{MT}.

This paper presents a new fast digital algorithm for the \ac{MT} that is both of low computational complexity and robust to noise present in projection data. The algorithm, which is denoted as the \ac{FMT} for convenience, symmetrises the \ac{MT} and its projection set to simplify its inversion. A novel analytical and exact mapping of Mojette projections to the \ac{DFT} is also constructed. This mapping, together with a new Mojette set definition, allows the direct and exact ``reformatting'' of Mojette projections so as to completely tile all of its corresponding Discrete Fourier space. The set, which ranges from $N+1$ to $N+\nicefrac{N}{2}$ number of projections for an $N\times N$ image, completely fills \ac{DFT} space without interpolation and filtering, thus allowing an exact reconstruction. If any filtering is required, it is exact and does not introduce any artefacts or approximations in the reconstruction. The issues associated with detector geometry and resolution of the Mojette projections (as given by \eqnTag~\eqref{eqn::MT_BinNumber}) are also addressed. The result is that the projection angles are not evenly spaced, the acquisition of projections takes advantage of the digital nature of the reconstruction, the inversion is fast and there are no reconstruction artefacts. Section~\ref{sec::Acquisition} proposes an experimental set-up for utilising this \ac{FMT}, while preserving its simple inversion process. The \ac{FMT} is made possible by utilising another discrete \ac{RT} known as the \ac{DRT}.

\subsection{Discrete Radon Transform}
The \acf{DRT}, which was the first discrete \ac{RT} constructed~\citep{Jordan1870, Kung1979, Labunets1985, Diaconis1985, Grigoryan1986, Bolker1987, Gertner1988, Fill1989, MatusFlusser}, is the \ac{RT} constructed within the same finite geometry as the \ac{DFT}, i.e. the image is assumed to possess periodic boundaries. The \ac{DRT} provides an exact partition of \ac{DFT} space in the form of periodic slices, where a periodic slice is the \ac{1D} \ac{DFT} of a \ac{DRT} projection. The partitioning is known as the Discrete \ac{FST}~\citep{Grigoryan1986, MatusFlusser}. The fast version of the \ac{DRT}, that utilises the \ac{FFT}~\citep{Cooley1965, Rader1968}, is denoted as the \ac{FRT} in what follows.

\begin{figure}[\placementHere]
 \centering
 \subfloat[]{
 \includegraphics[width=0.45\textwidth]{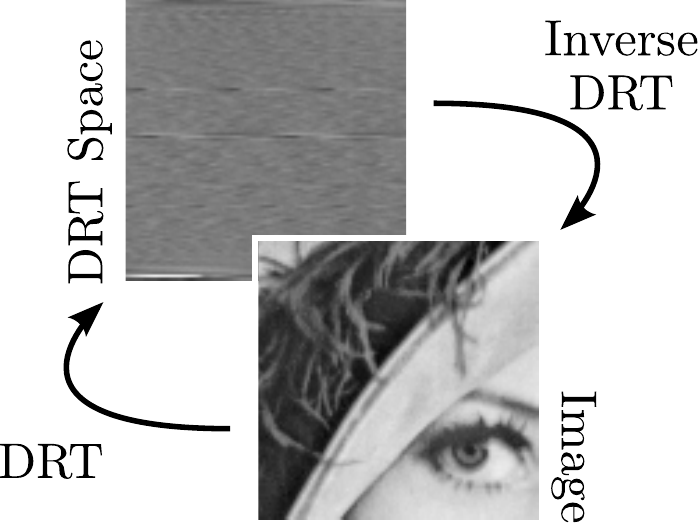}}
 \hspace{0.5cm}
 \subfloat[]{
 \includegraphics[width=0.4\textwidth]{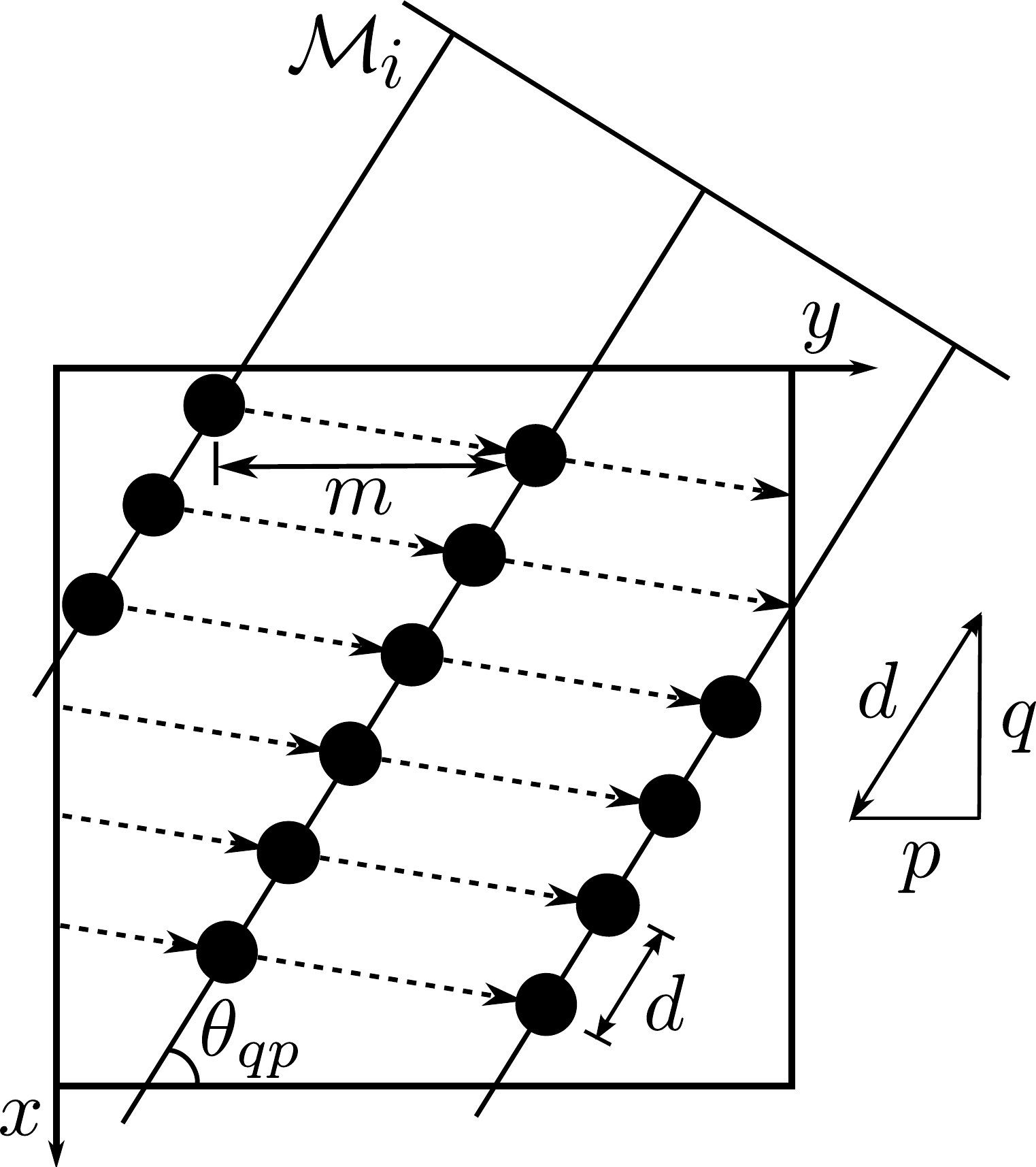}}
 \caption{The \acl*{DRT}~(\acs*{DRT}). (a) shows the exact or information preserving nature of the \acs*{DRT}. (b) shows the projections within the \acs*{DRT} (dotted lines) and how they are numerically mapped to linogram (\acl*{MT}) projections $\transformMojette_i$ (solid lines) using the nearest neighbour distance $d$ between pixels sampled.}
 \label{fig::FRT}
\end{figure}

\begin{figure}[\placement]
 \newcommand{\samplespacewidth}{0.4\textwidth}
 \newcommand{\samplehspace}{0.25cm}

 \centering
 \subfloat[]{
 \includegraphics[width=\samplespacewidth]{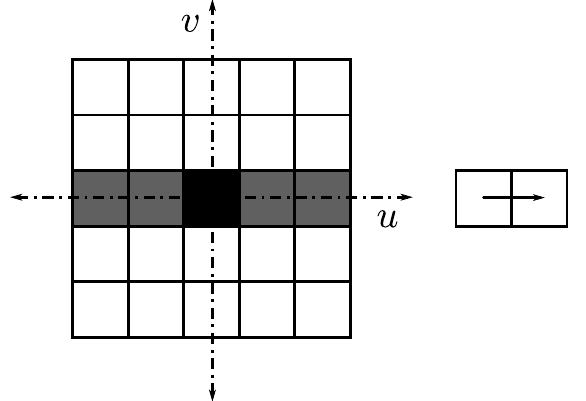}}
 \hspace{\samplehspace}
 \subfloat[]{
 \includegraphics[width=\samplespacewidth]{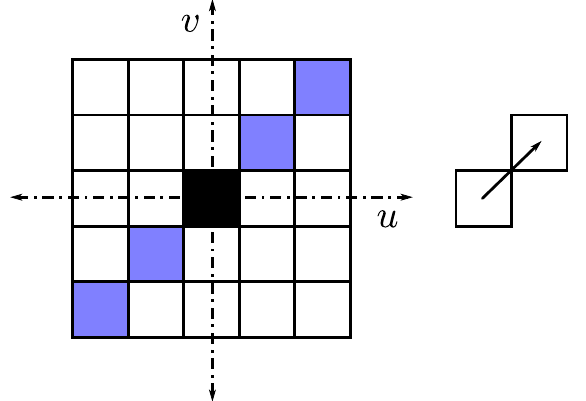}}
 
 \subfloat[]{
 \includegraphics[width=\samplespacewidth]{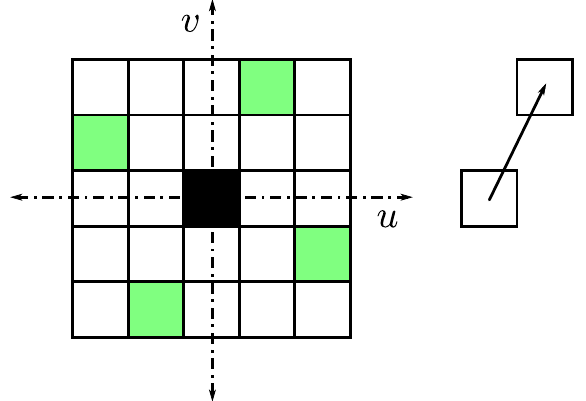}}
 \hspace{\samplehspace}
 \subfloat[]{
 \includegraphics[width=\samplespacewidth]{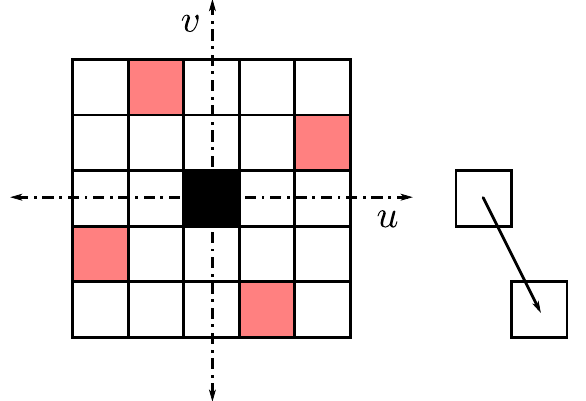}}
 
 \subfloat[]{
 \includegraphics[width=\samplespacewidth]{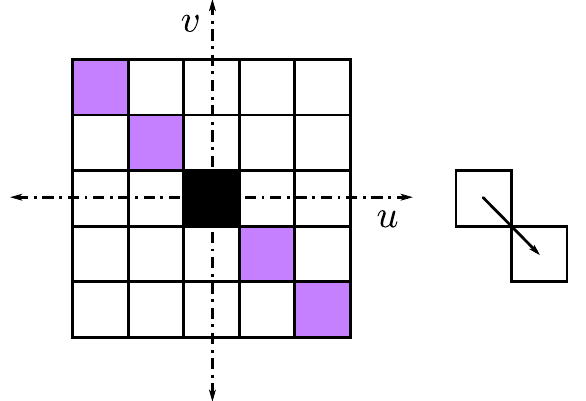}}
 \hspace{\samplehspace}
 \subfloat[]{
 \includegraphics[width=\samplespacewidth]{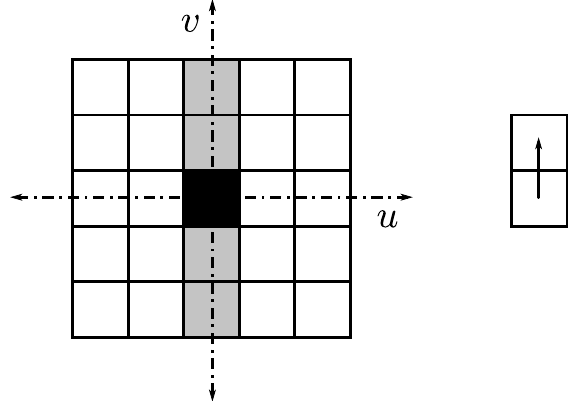}}

 \caption{The exact but periodic slices within the \acf*{FFT} (or equivalently the \acl*{FRT}) for the prime case $\primeSymbol=5$. Each colour represents a slice of a different slope with the DC coefficient/origin centred (black). Note that each vector shown is computed modulo $\primeSymbol$. (a)-(e) shows the slices with slopes $0 \leqslant m < 4 \imod 5$ in \acs*{FFT} space. (f) shows the row sum (perpendicular) slice in \acs*{FFT} space. The set of $\primeSymbol+1$ slices tile all of the space exactly once.}
 \label{fig::FST}
\end{figure}

The \ac{2D} Cartesian \ac{DFT} is tiled using \ac{1D} slices (see \figTag~\ref{fig::FST}) along the congruences (periodic lines)
\begin{eqnarray}
 \mat{t} &\equiv& y - \mat{m}x \imod N,\label{eqn::DiscreteLines1}\\
 \mat{t} &\equiv& x - \primeSymbol \mat{s}y \imod N,\label{eqn::DiscreteLines2}
\end{eqnarray}
for an $N\times N$ \ac{FFT} space with $N=\primeSymbol^n$ and $\primeSymbol$ is a prime number (hence, including powers of two)~\citep{Kingston2007}. The lines \eqref{eqn::DiscreteLines1} and~\eqref{eqn::DiscreteLines2} are utilised for the slopes
\begin{eqnarray}
 \mat{m} &=& \left\{m \setSeparator m < \numberSymbol,\ m\in\wholeNumbers \right\},\\
 \mat{s} &=& \left\{s \setSeparator s < \nicefrac{\numberSymbol}{\primeSymbol},\ s\in\wholeNumbers \right\},
 \label{eqn::DRTProjectionSlopes}
\end{eqnarray}
and the set of translates $\mat{t}$ as
\begin{equation}
 \mat{t} = \left\{t \setSeparator t < \numberSymbol,\ t\in\wholeNumbers \right\},
\end{equation}
so that the projections are effectively acquired along the vectors $[1,m]$ and $[\primeSymbol s,1]$ modulo $N$, while the slices are placed along the vectors $[-m,1]$ and $[1,-\primeSymbol s]$ in \ac{FFT} space (see \figTag~\ref{fig::FST}). Note that the \ac{FRT} projections/slices are all of the same size or length, unlike the \ac{MT}, which makes the \ac{FRT} symmetric in comparison and easy to invert (see \figTag~\ref{fig::FRT}(a)).

To completely tile all elements of the entire space at least once, i.e. cover all possible coefficients in the $N\times N$ \ac{FFT} space, a total of $N+\nicefrac{N}{\primeSymbol}$ slices (and hence projections) are required~\citep{Kingston2007}. For the simplest prime case $n=1$ so that $N+1$ projections are required, the $\gcd(m,\primeSymbol)=1$ always, so the slices of the same translate $t$ may only intersect once at the point $(0,t)$. The space is then tiled exactly and exact inversion is possible~\citep{MatusFlusser}. For the case $n>1$ so that $N+\nicefrac{N}{\primeSymbol}$ projections are required, those lines having slopes coprime to $\numberSymbol$ will intersect only once at the DC coefficient for a fixed $t$. However, the slopes not coprime to $\numberSymbol$ will intersect $\primeSymbol$ times, leading to under-sampling. The extra $\nicefrac{\numberSymbol}{\primeSymbol}$ projections perpendicular to $m$, i.e. at slopes $s$, are needed in order to sample all of the image. This results in a certain amount of oversampling, which is easily and exactly corrected by dividing each coefficient $u$ for all $N+\nicefrac{N}{\primeSymbol}$ slices by $\gcd(u,N)$~\citep{Kingston2007}.

Thus, for a $512\times 512$ object, one needs a total of $768$ slices for an exact reconstruction via the power of two \ac{FRT}. The same image would require $522$ slices when padded to the nearest prime number $521$ using a fast prime-sized \ac{FFT} via Rader's algorithm~\citep{Rader1968}, which is approximately two to three times slower than the power of two \ac{FFT} (see \figTag~\ref{fig::FST} for an example of a prime-sized tiling). The reconstruction is then simply obtained exactly by using the inverse \ac{FFT}. Reconstruction can be started as soon as slices become available, similar to conventional tomographic methods.

Chandra~\citep{Chandra2010c} also showed that the projections can be mapped to the \ac{NTT}, an integer-only transform analogous to the \ac{FFT} but with better performance for digital projections. The main issue with the \ac{FRT} is that the projections are taken along the periodic lines \eqref{eqn::DiscreteLines1} and~\eqref{eqn::DiscreteLines2}, such as the slices shown in \figTag~\ref{fig::FST}. This makes the experimental geometry not physically meaningful, limiting its practical application to tomography. Despite this, the \ac{FRT} has been successfully applied to tomography, but no algorithm has resulted that is suitably experimentally and computationally practical~\citep{Salzberg1999, SvalbeVanDerSpek, Chandra2008, Fayad2008}. The next section describes the first major result of the paper, the fusion of the \ac{MT} and the \ac{FRT} to produce a discrete \ac{RT} that is both experimentally and computationally viable for tomography.

\section{Fast Mojette Transform}
\begin{figure}[\placement]
 \centering
 \includegraphics[width=0.6\textwidth]{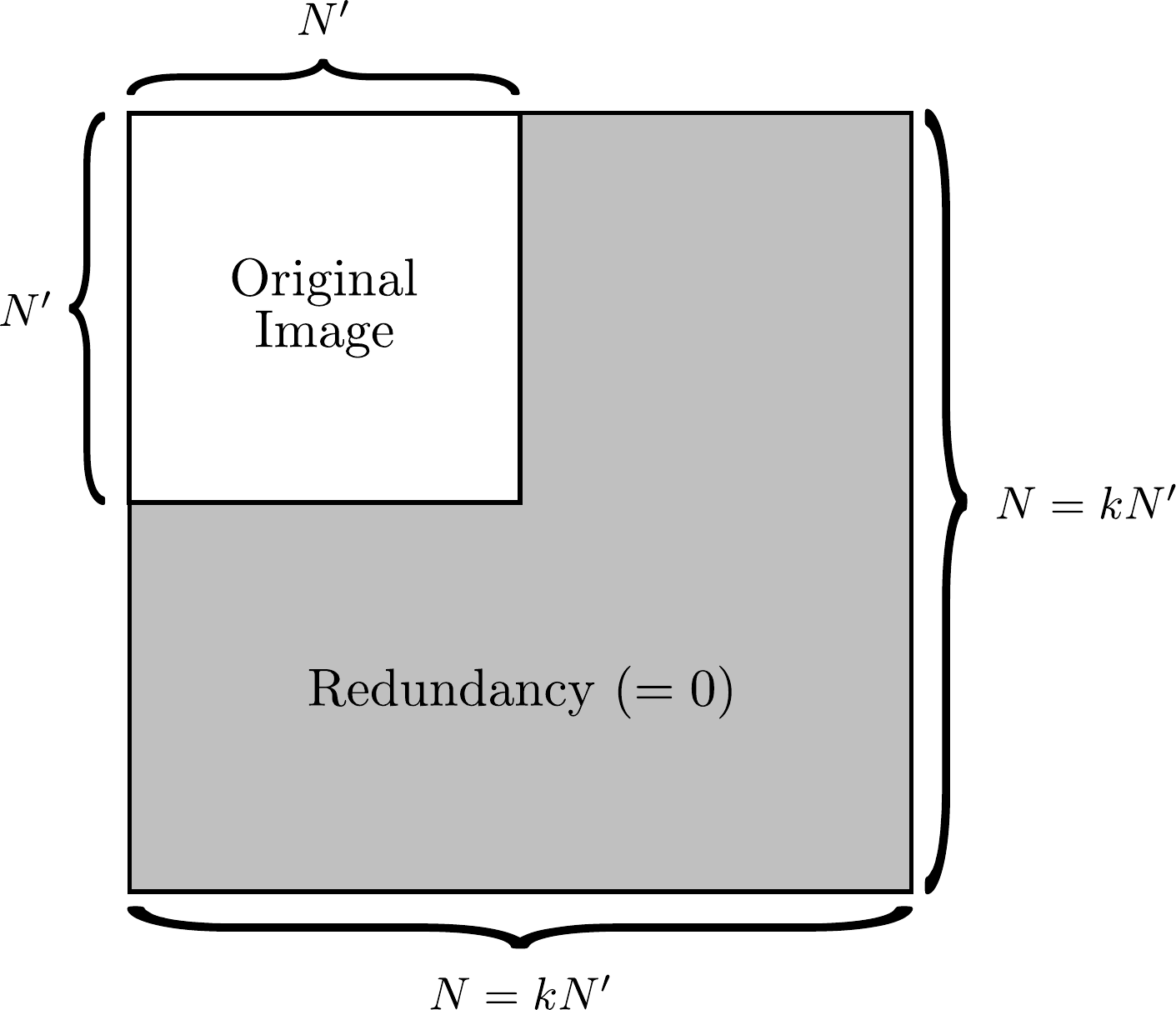}
 \caption{The redundancy within the \acl*{FMT} as an $N\times N$ image. The redundancy parameter $k$ is chosen so that $N = kN'$ is a power of two for a $N'\times N'$ image. The original image is assumed to be positioned at the top left for convenience.}
 \label{fig::Redundancy}
\end{figure}

In this section, the algorithm for fast and practical digital tomography is described. Given an $N'\times N'$ image, the objective is to exactly tile an $N\times N$ \ac{FFT} space, where $N = kN'$ and $k \geqslant 1$, so as to efficiently reconstruct the image with the inverse \ac{FFT} (i\ac{FFT}). For the \ac{FFT} space where $k > 1$, redundancy provides the mechanism for suppressing the effects of noise within the projections (see \figTag~\ref{fig::Redundancy}).

\begin{figure}[\placement]
 \centering
 \includegraphics[width=0.98\textwidth]{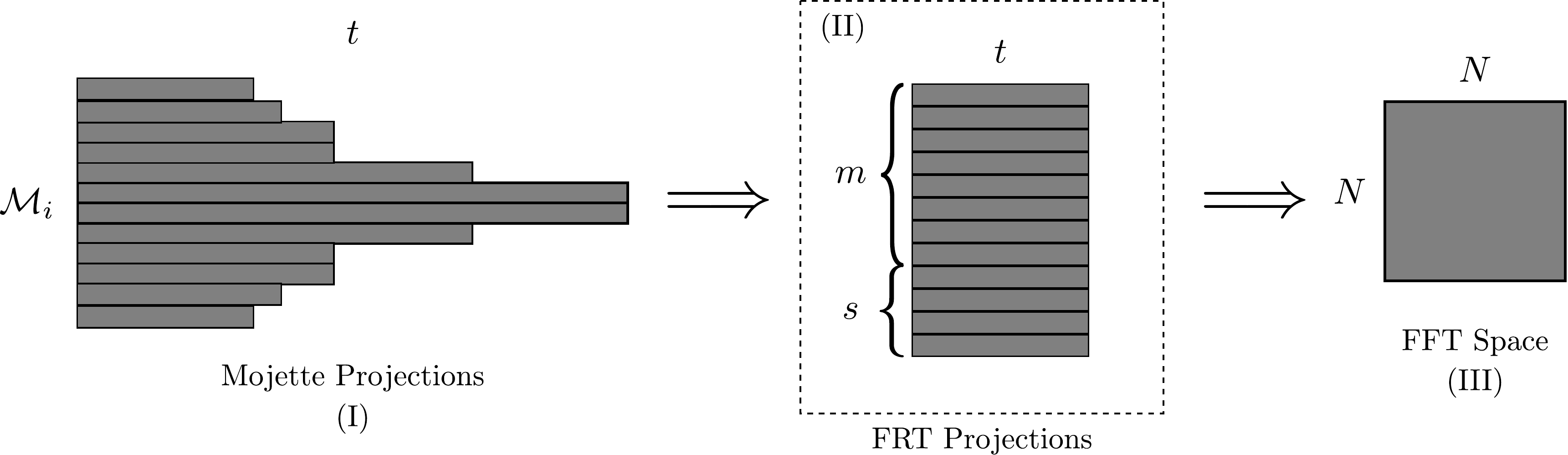}
 \caption{A schematic of the main processes within the \acl*{FMT} for $N=8$. The Mojette projections (I), acquired using a new Mojette set definition, are converted to \acl*{FRT} projections (II) via a new discrete projection mapping, and placed into \acl*{FFT} space (III) ready for inversion. The structure within the dashed box (II) is a logical step, i.e. the projections are converted in-place.}
 \label{fig::MT2FRT_Schematic}
\end{figure}

Assuming that the size of the image is $N=2^n$, in order to utilise the Cooley-Tukey~\citep{Cooley1965} algorithm for the \ac{FFT}, let the total number of projections $M = N+\nicefrac{N}{2}$. If each Mojette projection maps to a unique \ac{FRT} projection or slice, then they will tile all of \ac{FFT} space. Once the mapping is determined to be one-to-one, the Mojette projections must be converted to \ac{FRT} projections. A schematic of these processes are given in \figTag~\ref{fig::MT2FRT_Schematic}. 
It will be shown that both these key aspects are possible analytically and are exact in nature. A method is also given to generate optimal projection angle sets. It is then shown how one could experimentally acquire these projections and apply methods to control the noise response within the reconstructions.

\subsection{Finite Projection Mapping}
The fundamental operation for taking Mojette projections is the same as the \ac{FRT} because whole pixel intensities are summed along digital rays and placed into discrete bins, i.e. the \ac{DPM}. The main difference is periodicity, i.e. the \ac{MT} consist of line segments forming parts of the lines formed from the congruences \eqref{eqn::DiscreteLines1} and~\eqref{eqn::DiscreteLines2} of the \ac{FRT}. The segments can be linked using the shortest distance $d$ vectors between pixels sampled by these segments (see \figTag~\ref{fig::FRT}(b)). However, these vectors need to be computed using numerical methods which can be computationally inefficient~\citep{Svalbe2003}.

\begin{figure}[\placement]
 \centering
 \subfloat[]{
 \includegraphics[scale=0.9]{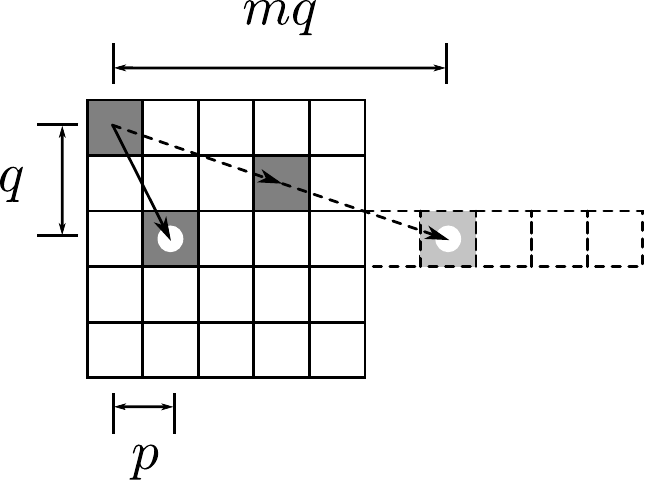}}
 \hspace{1cm}
 \subfloat[]{
 \includegraphics[scale=0.9]{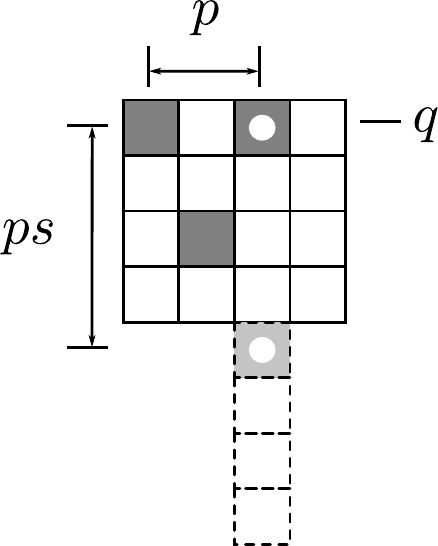}}
 \caption{A visual presentation of the mapping of \acl*{MT} projections to the \acl*{FRT}. Solving for $m$ in (a) and $s$ in (b) results in the mappings \eqref{eqn::mMap} and~\eqref{eqn::sMap}.}
 \label{fig::MT2FRT}
\end{figure}

The analytical mapping is as follows. For the case $n=1$ in $N=\primeSymbol^n$, the projection $\mu_{\theta_{pq}}$ of the \ac{MT} maps to the $m$ projection of the \ac{FRT} as
\begin{equation}
 m \equiv p q^{-1} \imod N,\label{eqn::mMap}
\end{equation}
where $q^{-1}$ is the multiplicative inverse of $q$ that can be computed easily via the Extended Euclidean algorithm. For the case $n>1$, the projection $\mu_{\theta_{pq}}$ maps to the $m$ and $s$ projections as \eqnTag~\eqref{eqn::mMap} when the $\gcd(p,N)>1$, and
\begin{equation}
 2s \equiv p^{-1} q \imod N,\label{eqn::sMap}
\end{equation}
when the $\gcd(q,N)>1$, respectively. These mappings cover all possible vectors/angles because the $\gcd(p,q)=1$ by definition of the angles. The mapping results from solving $mq \equiv p \imod N$ and $2sp \equiv q \imod N$ respectively. Examples of how each of these congruences are determined is shown in \figTag~\ref{fig::MT2FRT}.

It is crucial to ensure that the Mojette set has a one-to-one correspondence to the \ac{FRT} set, so that the \ac{FRT} (and equivalently \ac{FFT}) space is filled completely. This new Mojette set definition, which is finite in number and spans the all octants in the half plane, while providing a discrete coverage the interval $[0,\pi)$, is constructed in the next section.

\subsection{Angle Set}
\begin{figure}[\placement]
 \centering
 \subfloat[]{
 \includegraphics[width=0.49\textwidth]{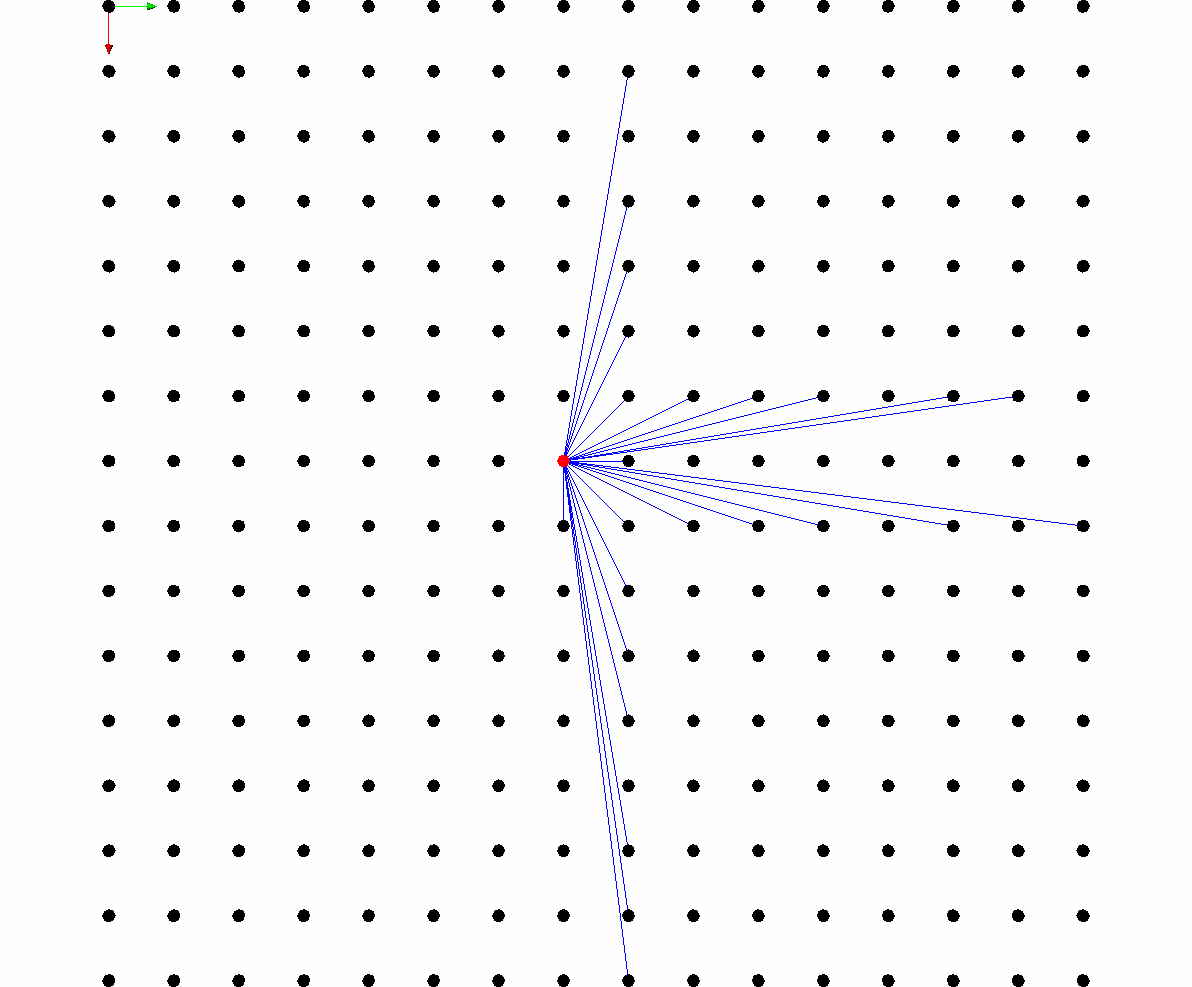}}
 \subfloat[]{
 \includegraphics[width=0.49\textwidth]{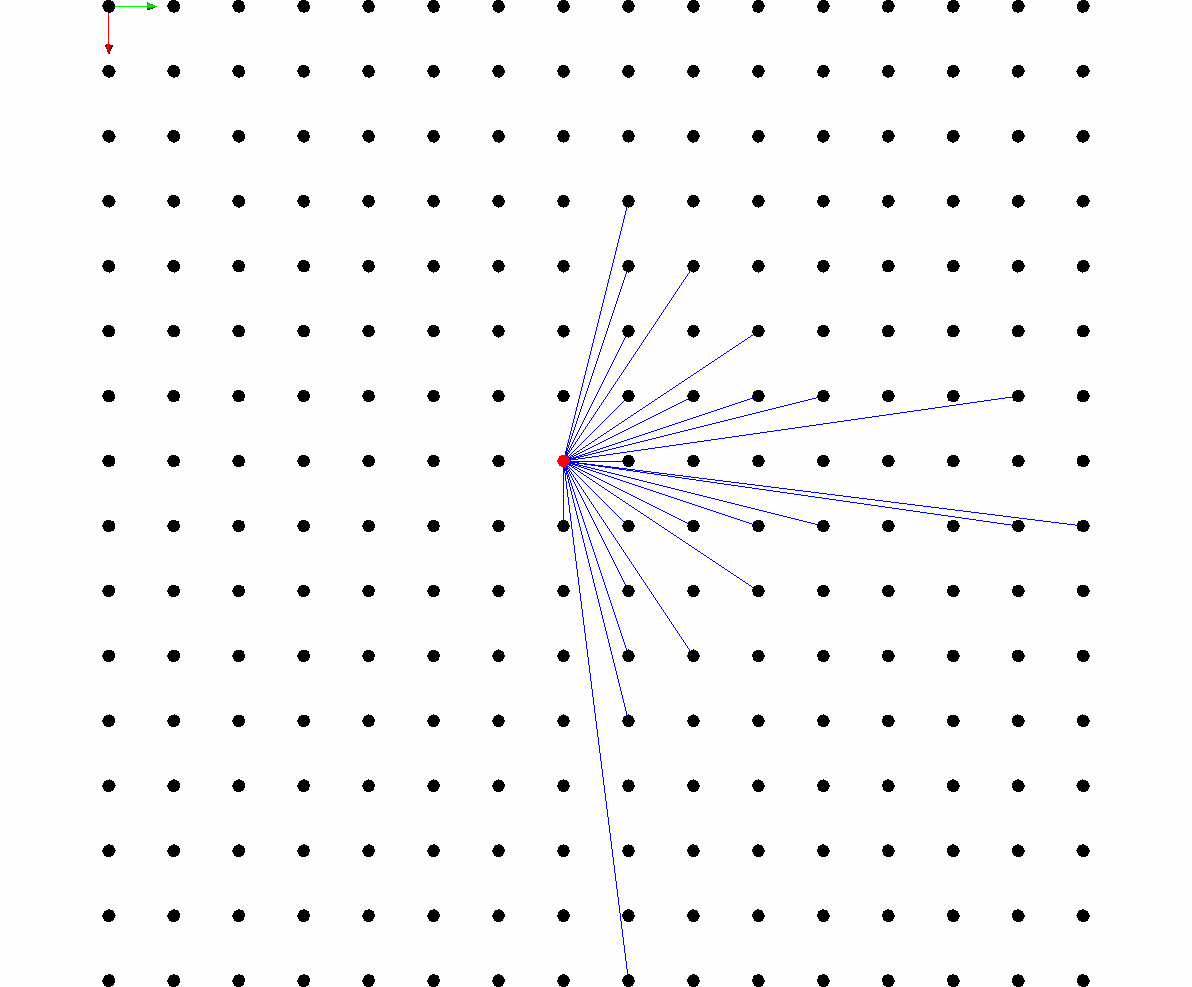}}
 \caption{(a) shows an example of a simple angle set, formed from the vectors $[1,a]$ with $a=0,\ldots,N-1$ in all octants in the half plane, used within the \acl*{FMT} for $8\times 8$ array. Note that this set is not necessarily symmetric in all octants. (b) shows an example of a true $\ell_1$ minimised angle set for the same array. Both mappings tile or cover \ac{FRT} (and hence \ac{FFT}) space for the image exactly. The origin is marked as a red point.}
 \label{fig::Angles}
\end{figure}

In this section, an \ac{FMT} set is constructed that optimally fills \ac{FRT} space completely and avoids Mojette projections that are ``long'', i.e. projections with a large number of translates $B$, by minimising \eqnTag~\eqref{eqn::MT_BinNumber}. This is equivalent to minimising the $\ell_1$-norm of the Farey fractions/vectors as
\begin{equation}\label{eqn::L1}
\ell_1 = |p|+|q|,
\end{equation}
for each projection in the Mojette set. The simplest way to achieve this is to generate angles through the vectors $[1,w]$ with $w\in\mathbb{Z}$ in all octants of the half plane. The $m$ or $s$ value is then checked with the \ac{FRT} set in question for each vector $[1,w]$ using the mappings \eqref{eqn::mMap} and~\eqref{eqn::sMap}. An example of this set for an $8\times 8$ image is given \figTag~\ref{fig::Angles}(a). Note that this set is not the same as the \ac{FRT} vectors $[1,m]$ and $[\primeSymbol s,1]$, but shares their simplicity. This angle set is a subset of the possible minimal $\ell_1$ vectors and is computationally inexpensive to generate. Hence, the set is well suited when adapting the detector or image geometry.

To generate the minimal $\ell_1$ angle set, one computes
\begin{equation}\label{eqn::CompactFarey}
q_3 = \left\lfloor\frac{q_1+p_1+N}{p_2}\right\rfloor q_2-q_1,\ \quad p_3 = \left\lfloor\frac{q_1+p_1+N}{p_2}\right\rfloor p_2-p_1,
\end{equation}
where $\lfloor\ \cdot\ \rfloor$ is the floor (round-down) operators, beginning the computation with $[q_1,p_1]=[0,0]$ and $[q_2,p_2] = [1,N]$ until $[q_3,p_3]=[1,1]$. \EqnTag~\eqref{eqn::CompactFarey} is a modified form of the standard way to generate Farey fractions. An example of this set for an $8\times 8$ image is given in \figTag~\ref{fig::Angles}(b). Once generated, the set must be sorted by ascending $\ell_1$ and the first $N+\nicefrac{N}{\primeSymbol}$ mappings to the \ac{FRT} angle set chosen. However, the set requires the generation of a large number of fractions and sorting of these fractions, making the computation expensive. Thus, this set is optimal for an unchanging detector and image geometry, since it need only be computed once.

\subsection{Finite Conversion}
Once the projection set is known for a given image and \ac{FFT} space, the Mojette projections need to be converted to the finite \ac{FRT} projections. This can be done by equating \eqnsTag~in~\eqref{eqn::DiracPM}, as well as \eqnsTag~\eqref{eqn::DiscreteLines1} and~\eqref{eqn::DiscreteLines2}, with the mappings \eqref{eqn::mMap} and~\eqref{eqn::sMap}.  For a Mojette translate $t_\transformMojette$ and an \ac{FRT} translate $t_\transformDiscreteRadon$, \eqnTag~\eqref{eqn::DiracPM} becomes
\begin{eqnarray*}
 q^{-1}\ t_\transformMojette &=& q^{-1} q y - p\ q^{-1} x\nonumber\\
 &=& y - p\ q^{-1} x,\nonumber
\end{eqnarray*}
when multiplying by $\ q^{-1}$. This expression is equivalent to \eqnTag~\eqref{eqn::DiscreteLines1} via \eqnTag~\eqref{eqn::mMap} when taken modulo $N$. Likewise, for the $s$ finite projections 
\begin{eqnarray*}
 p^{-1}\ t_\transformMojette &=& p^{-1} q y - p\ p^{-1} x\nonumber\\
 &=& p^{-1} q y - x,\nonumber
\end{eqnarray*}
which is equivalent to \eqnTag~\eqref{eqn::DiscreteLines2} via \eqnTag~\eqref{eqn::sMap} when taken modulo $N$. Similar equations result for the angles in the other octants. Hence,
\begin{eqnarray}\label{eqn::MT2FRT}
t_\transformDiscreteRadon = 
\begin{cases}
 q^{-1}\ t_\transformMojette \imod N, & \text{if}\quad \gcd(p,N) > 1\\
 p^{-1}\ t_\transformMojette \imod N, & \text{if}\quad \gcd(q,N) > 1\\
\end{cases}.
\end{eqnarray}
The conversion can be done prior to the inversion of the Mojette projections, after having been acquired and stored, or as the projections are being acquired. The latter is more desirable as the Mojette projections are effectively ``compacted'', since the \ac{FRT} projections are generally shorter than the Mojette projections and all have the same size. The next section discusses experimental methods to acquire Mojette projections.

\subsection{Mojette Acquisition}\label{sec::Acquisition}
Computationally, the Mojette projections can be acquired trivially using the \eqnsTag~in~\eqref{eqn::DiracPM}, since each pixel $(x,y)$ can be summed to its respective translate $t$~\citep{Guedon1995}. The major result of this paper is the experimental data acquisition geometry, which is the topic of the remainder of this section.

\begin{figure}[\placement]
 \centering
 \subfloat[]{
 \includegraphics[width=0.75\textwidth]{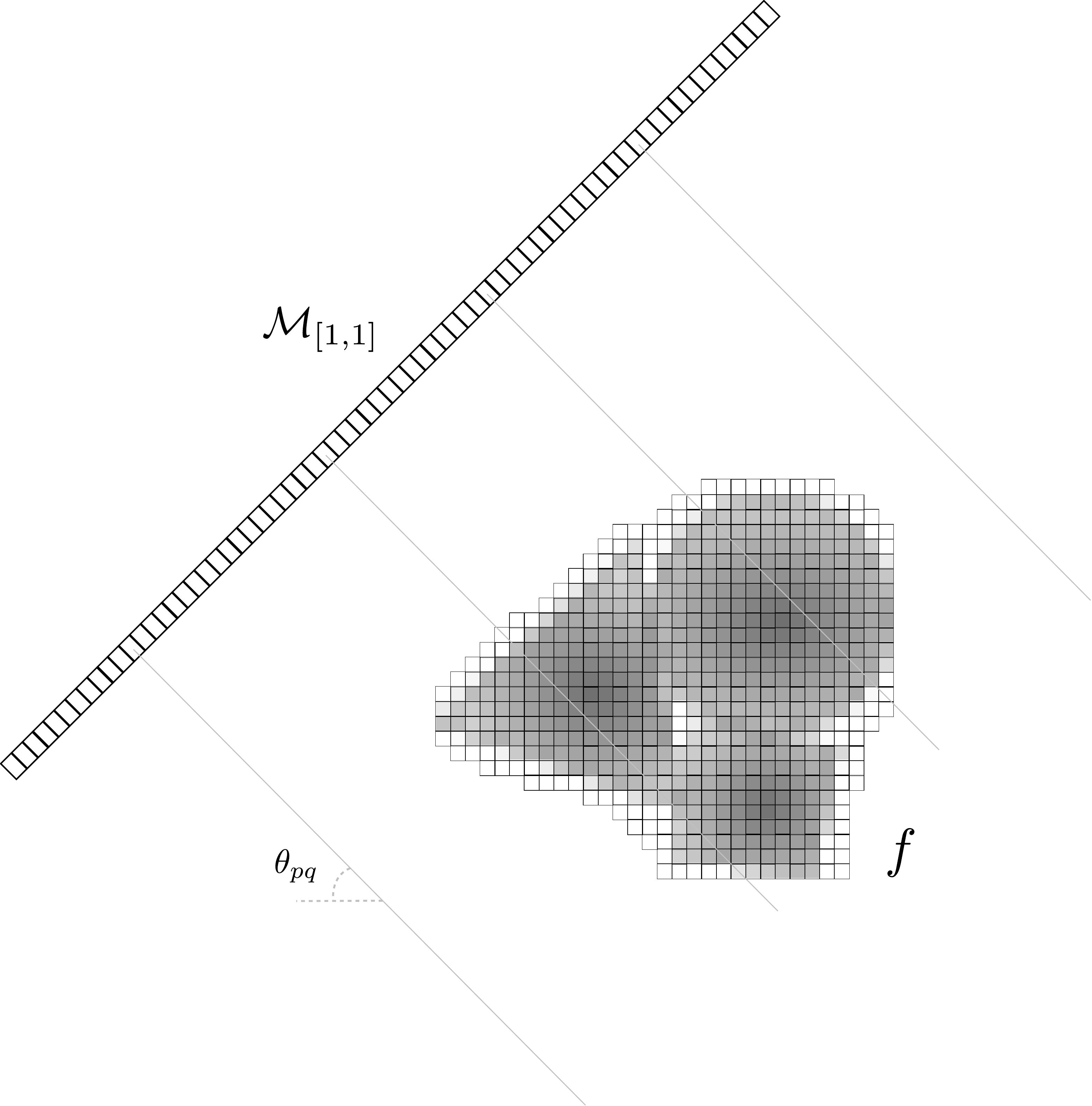}}

 \subfloat[]{
 \includegraphics[width=0.88\textwidth]{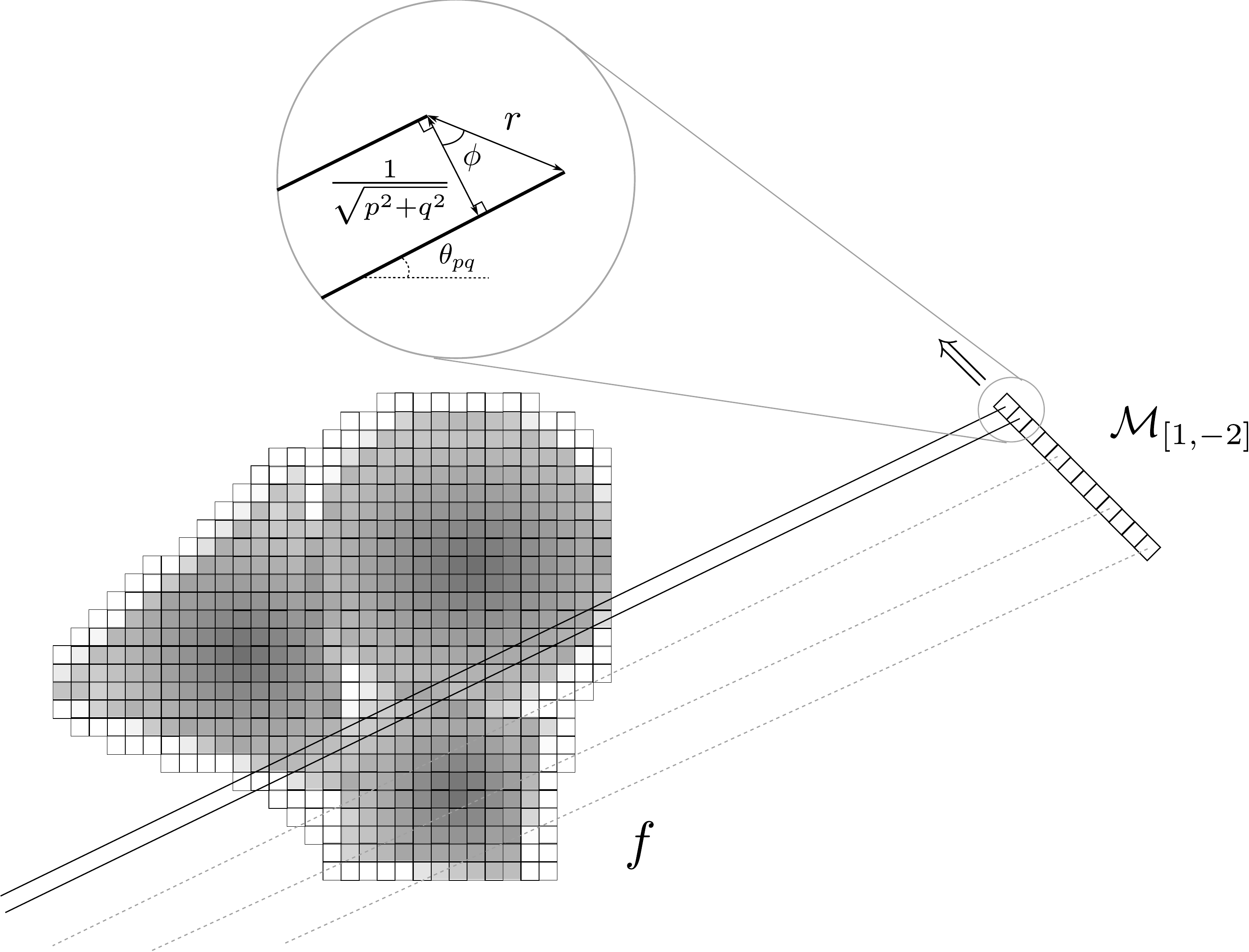}}
 \caption{Illustrations of the proposed experimental set-ups for the \acl*{FMT}. Experiment (a) utilises a large detector with a resolution equal to that of the Mojette projection requiring the highest resolution. Experiment (b) utilises a small fixed resolution detector that is scanned along at an angle $\phi$ (see inset and \eqnTag~\eqref{eqn::Phi}) with respect to the view angle of the Mojette projection. The angle $\phi$ is used to ensure that the (fixed width) pixels of the detector encompass the Mojette ray sums.}
 \label{fig::Exp}
\end{figure}

The \ac{MT} can be acquired experimentally in one of three ways. The simplest method is to use a \ac{1D} detector array of size $B_{\text{max}}$, which is typically at most $N^2$ in size for rectangular array $\Lambda$ when using a minimal $\ell_1$ angle set (see \figTag~\ref{fig::Exp}(a)).
This detector has to have the maximum resolution $r$ of 
\begin{equation}
 r = \frac{1}{\sqrt{p^2+q^2}},
\end{equation}
for unit pixel area, which is effectively the separation between the translates $t$ of the angle $\theta_{pq}$ corresponding to $B_{\text{max}}$ (see \citep[pg. 20-21]{Olds2000} for proofs). Since the remaining projections require a lower detector resolution, these projections become a sub-region within this detector and can be recovered by sub-sampling measurements on the detector. The other redundant parts of the detector in these cases may be used to estimate noise on those projections. The projections can then be converted exactly to \ac{FRT} projections using the \eqnsTag~in~\eqref{eqn::MT2FRT} and reconstruction process may be started immediately as slices become available.

For detectors smaller than $N^2$ in size, the detector can be translated or scanned across the object in the direction perpendicular to the projection angle, similar to the linogram approach~\citep{Edholm1988} (see \figTag~\ref{fig::Exp}(b)). To overcome limitations in fixed detector resolution and fixed pixel width or size $r$, the detector is scanned at angle $\nicefrac{\pi}{2}+\phi$ with respect to $\theta_{pq}$. The angle $\phi$ is given as
\begin{equation}\label{eqn::Phi}
 \phi = \cos^{-1}\left(\frac{1}{r\sqrt{p^2+q^2}}\right),
\end{equation}
for each vector $[q,p]$, where $\nicefrac{1}{\sqrt{p^2+q^2}}$ is the separation between the translates $t$ (see also inset of \figTag~\ref{fig::Exp}(b)). This ensures that the entire width of the pixel in the detector encompasses its corresponding ray sum without the need for a multi-resolution detector. However, the detector sensitivity may be adversely affected by the angle $\phi$.

The final method is to use a multi-resolution detector, either of size $B_{\text{max}}$ (being $N^2$ in size) or smaller to scan the detector at the angle $\nicefrac{\pi}{2}+\theta_{pq}$ in relation to the object. When utilising the \ac{DPM} for a continuous object, the projection needs to be convolved with a beam-spread function to account for fractional sampling of the array $\Lambda$~\citep{Kingston2003b}. For the simple angle set, this function is always a triangle with a base of $2w$ and a height of one for the vector $[1,w]$. In the next section, the issues regarding the suppression and the effects of noise are discussed.


\subsection{Noise}
The response of the \ac{FMT} algorithm to inconsistencies in projection data, such as noise from detectors, can be minimised in two non-exclusive ways. Firstly, the redundancy parameter $k$ can be used to ``spread out'' these inconsistencies over a larger image area. For noise that is statistical in nature, and therefore the same on average for different image sizes, the same noise is present over a larger area for larger image sizes. Thus, this noise is not as prominent over the reconstruction sub-region. Equivalently, the noise is averaged out due to having more translates in the projection data. However, since the larger image area requires more projections, it exposes the object to potentially more radiation. This is a non-issue for data encoding and/or transmission applications.

Secondly, the redundancy in the entire image, after the inversion of the \ac{MT}, can be used as part of a iterative algorithm for reducing the noise. For example, an $\ell_1$ or $\ell_2$-norm minimising (iterative) algorithm (applied to the pixel values) can be used to make image values in the redundant region converge to zero~\citep{Bube1997, Candes2006}. The result of this minimising algorithm gives an approximation to the noise in the projection data, reducing the effects of these inconsistencies within the image. This approach will be most useful in very noisy or incomplete projection data. 

The redundancy within the image area can also be utilised to reduce the number of projections as described by Chandra \emph{et al.}~\citep{Chandra2008} for the noise-free projection data. Any missing \ac{FRT} projections cause finite ghosts superimposed on the image that can be removed exactly via a non-iterative algorithm applied to these ghosts in the redundant area. The number of projections is then just the number of rows in the initial object, which is generally much less than $N+\nicefrac{N}{\primeSymbol}$. Ghost removal in the presence of noise is still an active area of research. The next section describes the performance and noise response of the \ac{FMT} and the \ac{MT}.

\subsection{Results}\label{sec::Results}
\begin{figure}[\placementHere]
 \centering
 \begin{minipage}{0.3\textwidth}
  \subfloat[]{
  \includegraphics[width=0.9\textwidth]{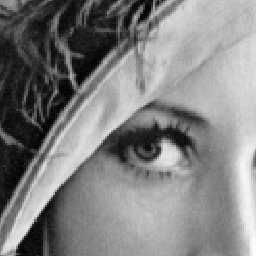}}
  \vspace{0.35cm}
  
  \subfloat[]{
  \includegraphics[width=0.9\textwidth]{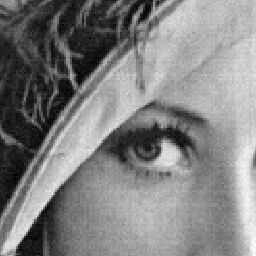}}
 \end{minipage}
 \hspace{0.15cm}
 \begin{minipage}{0.67\textwidth}
 \subfloat[]{
 \includegraphics[width=0.98\textwidth]{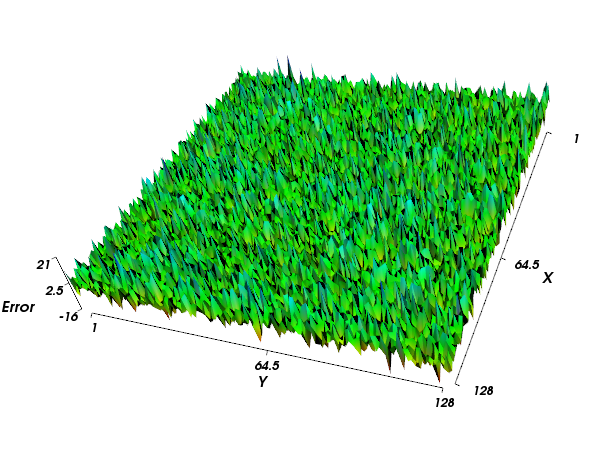}}
 \end{minipage}
 \caption{A noise response numerical simulation of the inverse \acl*{FMT} using the simple angle set. (a) shows the original $128\times 128$ image of Lena. (b) shows the reconstruction with $k=2$ and $3\%$ Gaussian noise. (c) shows the actual errors between (a) and (b) as a surface plot. On average, the error is approximately 5 out of 256 grey-scales per pixel.}
 \label{fig::Performance}
\end{figure}

Numerical simulations of the \ac{FMT} were conducted using the \ac{DPM} of a $128\times 128$ image of Lena on a Intel\texttrademark~Core2Duo\texttrademark~E6600 (2.4GHz) processor. \figTag~\ref{fig::Performance} shows an example of a simulated reconstruction using the \ac{FMT} of the image with 3\% Gaussian noise present in the projections within a $256\times 256$ \ac{FFT} space (i.e. the redundancy parameter of $k=2$).
The reconstruction (\figTag~\ref{fig::Performance}(b)) shows that the result is stable to moderate levels of noise, having a Peak-Signal-to-Noise (PSNR) of approximately 35dB, possibly suitable for lossy image and video encoding. The errors present on the reconstruction are shown in \figTag~\ref{fig::Performance}(c), which has a Root Mean Squared Error (RMSE) of 4.8 grey-scales out of a possible 8-bit (256) grey-scales. Note that the noise is uniformly distributed and no image related artefacts are apparent.
\begin{figure}[\placementHere]
 \centering
 \subfloat[]{
 \includegraphics[width=0.9\textwidth]{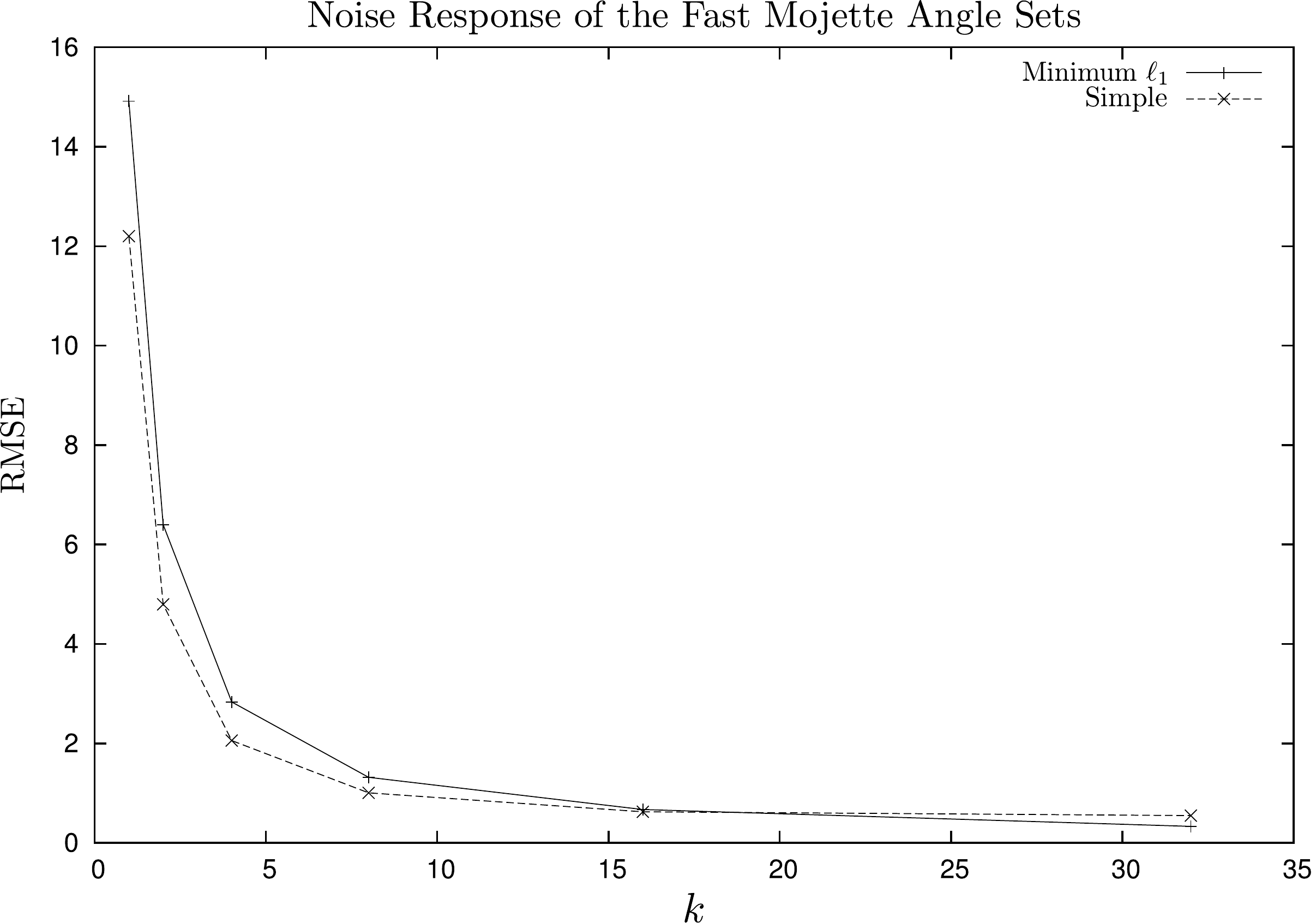}}
 
 \subfloat[]{
 \includegraphics[width=0.9\textwidth]{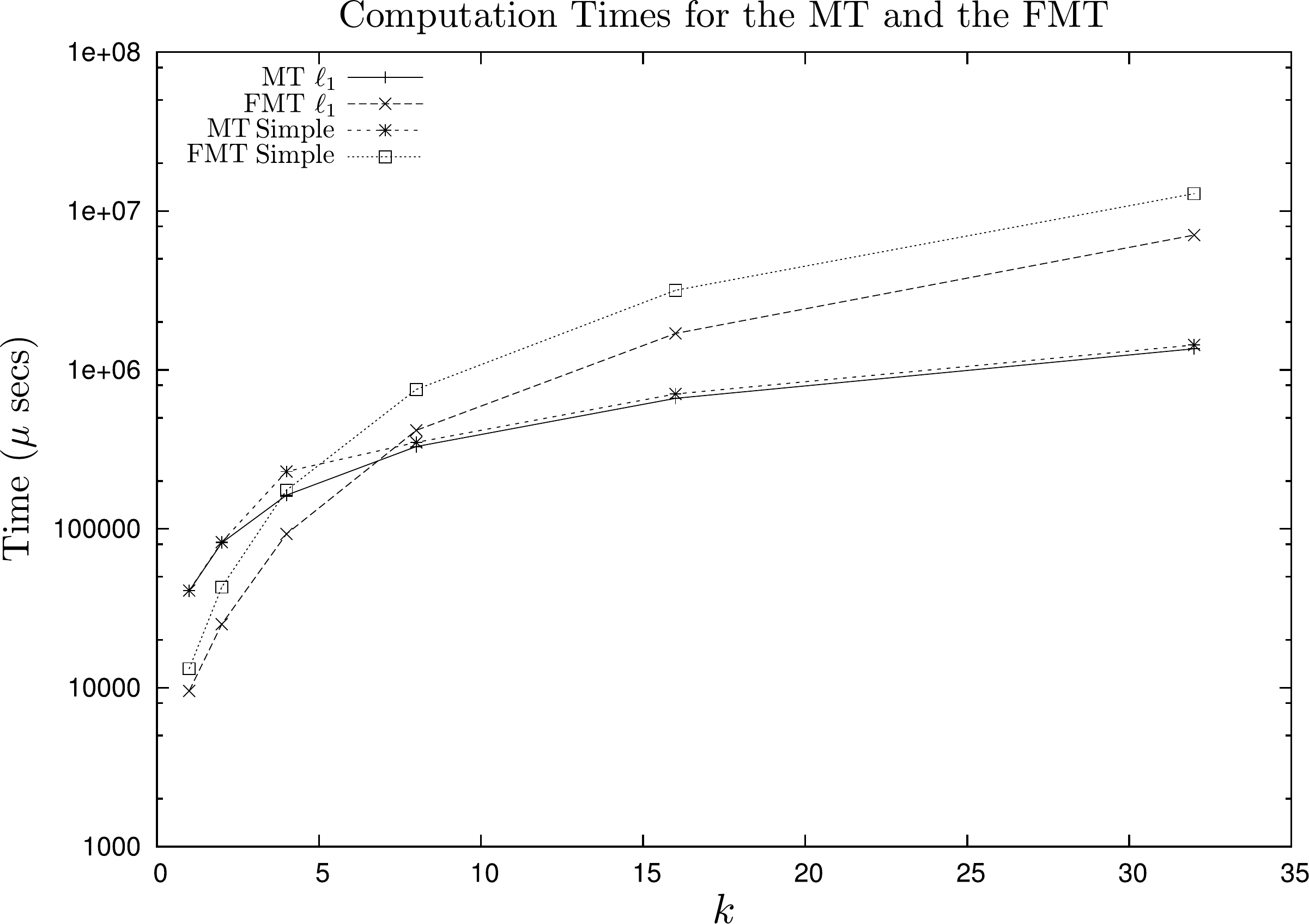}}
 \caption{A comparison of the angle sets for the \acf*{FMT} with a $128\times 128$ image of Lena. (a) shows the Root Mean Squared Error (RMSE) with increasing redundancy $k$. (b) shows the computation times of the inverse \ac*{FMT} and the forward \acf*{MT} on a log scale with the time in micro-seconds ($\mu$ secs).}
 \label{fig::Graphs}
\end{figure}

Graph~\ref{fig::Graphs}(a) shows the noise response of the two different angle sets as a function of the redundancy parameter $k$. The parameter values $k$ are defined as powers of two to ensure that the \ac{FFT} space is also a power of two for the Cooley-Tukey algorithm~\citep{Cooley1965}. Increasing the parameter $k$ spreads out the noise as predicted and exponentially decreases the error on the reconstruction. The simple angle set, which is a subset of the $\ell_1$ minimum set, has a better noise response because it possesses more translates than the full $\ell_1$ minimum set.

Graph~\ref{fig::Graphs}(b) shows the computation times of the \ac{MT} and the inverse \ac{FMT} in micro-seconds. The inverse \ac{FMT}, although an order of magnitude slower than the \ac{MT} for very large images, is suitably fast as the time for reconstructing a $4096\times 4096$ image from 6144 projections is approximately 13 seconds. The time could be greatly reduced using Graphical Processing Units (GPUs). Considering both graphs, the optimal value for the parameter $k$ appears to be around $k=2$ or 4, in order to balance noise response with the speed of computation. Further work on the \ac{FMT} needs to be done in selecting optimal experimental geometries and conducting tomographic experiments with real data.

\section*{Conclusion}
A noise tolerant algorithm for fast digital-to-digital tomography was constructed (see graphs in \figTag~\ref{fig::Graphs}). The angle set for the \acl{MT} was redefined to one that was more symmetric and minimal in size, so that inversion could be computed with low complexity (see \eqnTag~\eqref{eqn::L1}). Two choices for the angle sets were constructed that were either computationally or experimentally suitable for variable or fixed image geometry respectively (see \eqnsTag~\eqref{eqn::mMap} and~\eqref{eqn::sMap}). A new analytical mapping of these projections was constructed that allowed them to be compacted directly and exactly into a Discrete Fourier space of desired size (see \eqnTag~\ref{eqn::MT2FRT}). Once the projections are mapped, the (robust to noise) reconstruction can be obtained with a computational complexity $O(n\log_2 n)$ with $n = N^2$ for an $N\times N$ space (see \figTag~\ref{fig::Performance}). Redundancy within this space was used to control inconsistencies in projection data, such as detector noise. The redundancy also allows the exploitation of $\ell_1$ or $\ell_2$-norm minimising (iterative) algorithms to values within the space to further reduce noise. Experiments and techniques to apply this \acl{FMT} for real measurements were also proposed (see \figTag~\ref{fig::Exp}).

\section*{Acknowledgements}
S. Chandra would like to thank the Faculty of Science, Monash University for a Ph.D scholarship and a publications award. N. Normand would like to thank the Australian Research Council for his International Fellowship.

\appendix
\section{Discrete Tomography Review}\label{sec::DRTs}
Discrete reconstruction methods take advantage of digital image geometry of the reconstructions to allow exact inversion schemes. The methods of Beylkin~\citep{Beylkin} and, Kelley and Madisetti~\citep{Kelley1992} utilise projections along arbitrary curves and reconstruction via Block-Circulant matrices. However, the method requires the storage of a large number of matrices and has an unfavourable computational complexity of $O(n^3)$ for a \ac{2D} image.
\begin{figure}[\placement]
 \centering
 \includegraphics[width=0.9\textwidth]{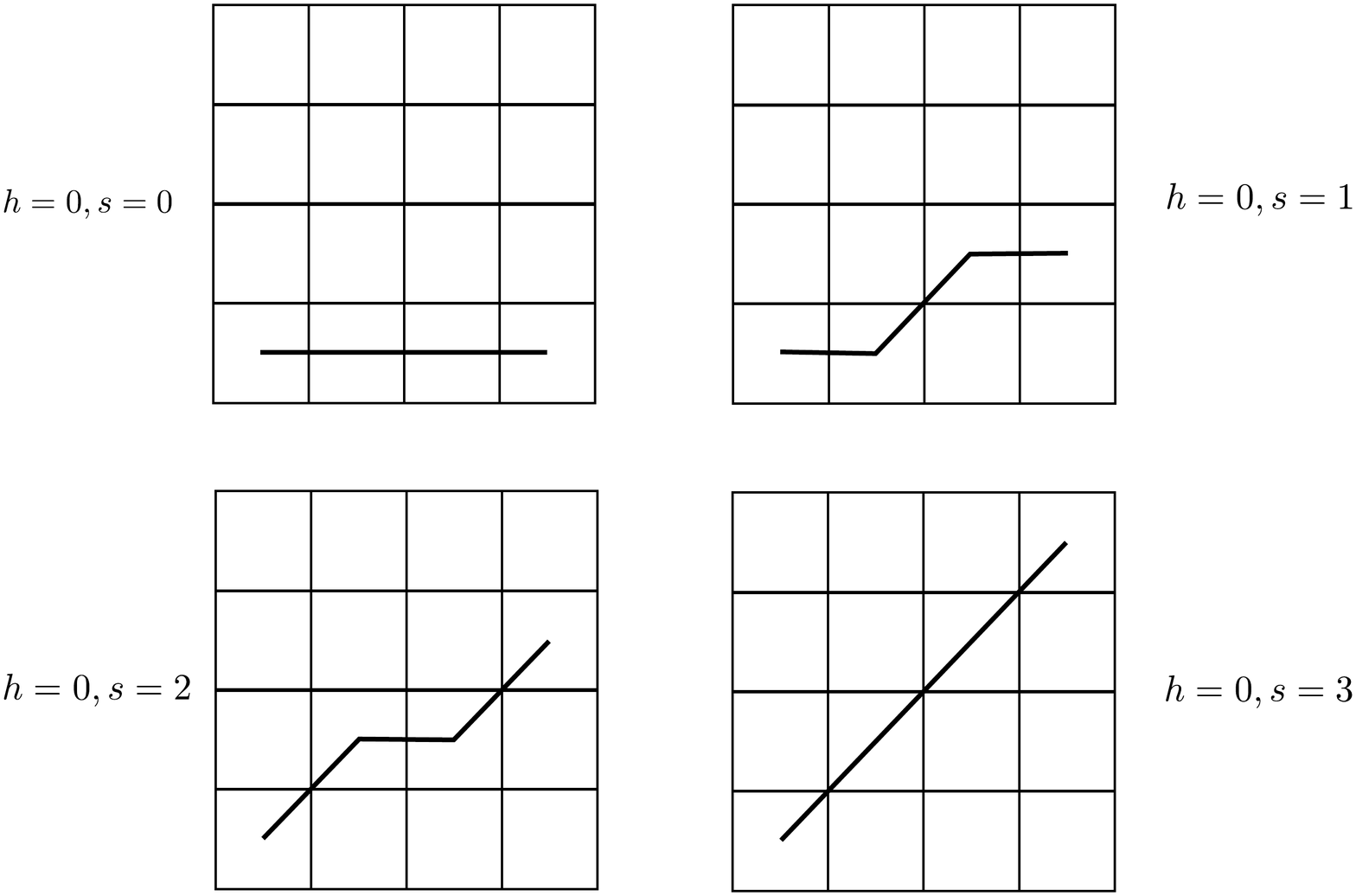}
 \caption{The $d$-lines for a $P\times Q$ image used to approximate continuous lines~\citep{Gotz1996,Brady1998}. A line, parametrised by $x$-intercept $h$ and rise $s$, is divided into two lines comprising the $0$ to $P/2-1$ and $P/2$ to $P-1$ pixels respectively. The final $y$-intercept of the first becomes $h+\lfloor\frac{s}{2}\rfloor$ and the $x$-intercept for the second is redefined as $h+\lfloor\frac{s+1}{2}\rfloor$.}
 \label{fig::d-lines}
\end{figure}

G\"otz and Druckm\"uller~\citep{Gotz1996}, and Brady~\citep{Brady1998} independently introduced the concept of $d$-lines, which approximate lines of arbitrary (including irrational) slopes (see \figTag~\ref{fig::d-lines}). Sums along $d$-lines approximate the classical linogram and an iterative algorithm has been developed which can recover the image with desired accuracy by Press~\citep{Press2006}.
\begin{figure}[\placement]
 \centering
 \includegraphics[width=0.9\textwidth]{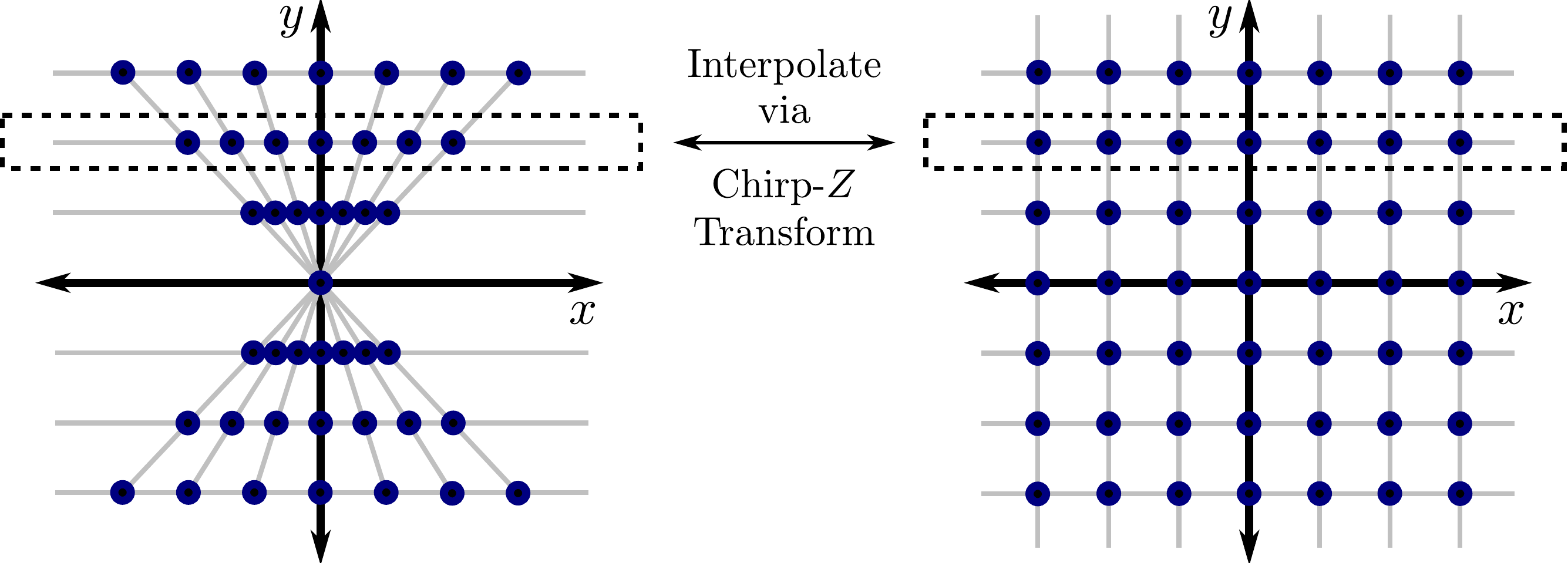}
 \caption{The \acl*{FST} mapping within the Fast Slant Stack~\citep{Averbuch2001}. The row-wise pseudo-polar sampling, created by the linogram mapping of slices, is cast to Cartesian Fourier space in order to use the \acl*{FFT}. The same is done to the columns of the mapping.}
 \label{fig::SlantStack}
\end{figure}

The last approach is to utilise the pseudo-polar \ac{FT}~\citep{AverbuchEtAl}. The Fast Slant Stack method bypasses the need for the filtering within the \ac{FST} via this pseudo-polar \ac{FT} having a linogram mapping of the slices~\citep{Averbuch2001}. However, the coefficients are still interpolated (via the Chirp-$Z$ transform) to Cartesian Fourier space in order to use the \ac{FFT} (see \figTag~\ref{fig::SlantStack}).

\small
\bibliographystyle{elsarticle-num}
\bibliography{RadonJabRef}

\begin{thebibliography}{10}
\expandafter\ifx\csname url\endcsname\relax
  \def\url#1{\texttt{#1}}\fi
\expandafter\ifx\csname urlprefix\endcsname\relax\def\urlprefix{URL }\fi
\expandafter\ifx\csname href\endcsname\relax
  \def\href#1#2{#2} \def\path#1{#1}\fi

\bibitem{RammKatsevich}
A.~G. Ramm, A.~I. Katsevich, {R}adon Transform and Local Tomography, CRC Press,
  1996.

\bibitem{Radon}
J.~Radon, \"{U}ber die bestimmung von funktionen durch ihre integralwerte
  l\"{a}ngs gewisser mannigfaltigkeiten, Berichte S\"{a}chsische Acadamie der
  Wissenschaften, Leipzig, Math.-Phys. Kl. 69 (1917) 262--267.

\bibitem{Funk1915}
P.~Funk, \"{U}ber eine geometrische {A}nwendung der {A}belschen
  {I}ntegralgleichung, Mathematische Annalen 77 (1915) 129--135.
\newblock \href {http://dx.doi.org/10.1007/BF01456824}
  {\path{doi:10.1007/BF01456824}}.

\bibitem{Smith1977}
K.~T. Smith, D.~C. Solmon, S.~L. Wagner, Practical and mathematical aspects of
  the problem of reconstructing objects from radiographs, Bulletin of the
  American Mathematical Society 83~(6) (1977) 1227--1270.

\bibitem{Logan1975}
B.~F. Logan, The uncertainty principle in reconstructing functions from
  projections, Duke Mathematical Journal 42~(4) (1975) 661--706.

\bibitem{Katz}
M.~Katz, Questions of Uniqueness and Resolution in Reconstruction from
  Projections, Lecture Notes in Biomathematics, Springer-Verlag, 1977.

\bibitem{Bracewell1954}
R.~N. Bracewell, J.~A. Roberts, {Aerial Smoothing in Radio Astronomy},
  Australian Journal of Physics 7 (1954) 615--640.

\bibitem{Louis1981}
A.~K. Louis, Ghosts in tomography - the null space of the {R}adon transform,
  Mathematical Methods in the Applied Sciences 3 (1981) 1--10.
\newblock \href {http://dx.doi.org/10.1002/mma.1670030102}
  {\path{doi:10.1002/mma.1670030102}}.

\bibitem{Beylkin}
G.~Beylkin, Discrete {R}adon transform, Acoustics, Speech, Signal Processing,
  IEEE Transactions on 35~(2) (1987) 162--172.

\bibitem{Kelley1992}
B.~Kelley, V.~Madisetti,
  \href{http://dx.doi.org/10.1109/ICASSP.1992.226189}{The fast discrete {R}adon
  transform}, ICASSP-92: 1992 IEEE International Conference on Acoustics,
  Speech and Signal Processing (Cat. No.92CH3103-9) (1992) 409 -- 12.
\newline\urlprefix\url{http://dx.doi.org/10.1109/ICASSP.1992.226189}

\bibitem{Gotz1996}
W.~G\"otz, H.~Druckm\"uller,
  \href{http://dx.doi.org/10.1016/0031-3203(96)00015-5}{A fast digital {R}adon
  transform-an efficient means for evaluating the {H}ough transform}, Pattern
  Recognition 29~(4) (1996) 711--718.
\newline\urlprefix\url{http://dx.doi.org/10.1016/0031-3203(96)00015-5}

\bibitem{Brady1998}
M.~L. Brady, \href{http://link.aip.org/link/?SMJ/27/107/1}{A fast discrete
  approximation algorithm for the {R}adon transform}, SIAM Journal on Computing
  27~(1) (1998) 107--119.
\newblock \href {http://dx.doi.org/10.1137/S0097539793256673}
  {\path{doi:10.1137/S0097539793256673}}.
\newline\urlprefix\url{http://link.aip.org/link/?SMJ/27/107/1}

\bibitem{Averbuch2001}
A.~Averbuch, D.~Donoho, R.~Coifman, M.~Israeli, J.~Walden, Fast slant stack: A
  notion of {R}adon transform for data on a cartesian grid which is rapidly
  computable, algebraically exact, geometrically faithful and invertible, Tech.
  Report, Stanford University 11.

\bibitem{Guedon1995}
J.-P. Gu\'edon, D.~Barba, N.~Burger, Psychovisual image coding via an exact
  discrete {R}adon transform, Proc. of the SPIE - The International Society for
  Optical Engineering 2501 (1995) 562--572.

\bibitem{Franel1924}
J.~Franel, Les suites de farey et le probl\'eme des nombres premiers, Gottinger
  Nachr (1924) 191–201.

\bibitem{Landau1924}
E.~Landau, Bemerkungen zu der vorstehenden abhandlung von herrn franel,
  G\"ottinger Nachr (1924) 202--206.

\bibitem{Guedon1997}
J.-P.~V. Gu\'edon, N.~Normand,
  \href{http://link.aip.org/link/?PSI/3024/873/1}{{M}ojette transform:
  applications for image analysis and coding}, Visual Communications and Image
  Processing '97 3024~(1) (1997) 873--884.
\newline\urlprefix\url{http://link.aip.org/link/?PSI/3024/873/1}

\bibitem{Normand2006}
N.~Normand, A.~Kingston, P.~\'Evenou, A geometry driven reconstruction
  algorithm for the {M}ojette transform, in: Lecture Notes in Computer Science
  (LNCS), Vol. 4245, Springer Berlin / Heidelberg, 2006, pp. 122--133.

\bibitem{Servieres2005a}
M.~Servi\`eres, N.~Normand, J.-P. Gu\'edon, Y.~Bizais, The {M}ojette transform:
  Discrete angles for tomography, in: G.~Herman, A.~Kuba (Eds.), Proceedings of
  the Workshop on Discrete Tomography and its Applications, Vol.~20, Electronic
  Notes in Discrete Mathematics, 2005, pp. 587--606.

\bibitem{NormandEtAl}
N.~Normand, J.-P. Gu\'edon, O.~Philippe, D.~Barba, Controlled redundancy for
  image coding and high-speed transmission, Proc. of the SPIE - The
  International Society for Optical Engineering 2727 (1996) 1070--1081.

\bibitem{Servieres2005}
M.~Servi\`eres, J.~Idier, N.~Normand, J.-P. Gu\'edon,
  \href{http://dx.doi.org/10.1117/12.593399}{Conjugate gradient {M}ojette
  reconstruction}, Proc. of the SPIE - The International Society for Optical
  Engineering 5747~(1) (2005) 2067--2074.
\newline\urlprefix\url{http://dx.doi.org/10.1117/12.593399}

\bibitem{Philippe1997a}
O.~Philipp\'e, J.-P. Gu\'edon, F.~Terrien,
  \href{http://link.aip.org/link/?PSI/3024/1220/1}{{ATM} source-channel image
  coding}, Visual Communications and Image Processing 3024~(1) (1997)
  1220--1230.
\newblock \href {http://dx.doi.org/10.1117/12.263202}
  {\path{doi:10.1117/12.263202}}.
\newline\urlprefix\url{http://link.aip.org/link/?PSI/3024/1220/1}

\bibitem{Terrien1997}
F.~Terrien, J.-P.~V. Gu\'edon, O.~Philippe,
  \href{http://link.aip.org/link/?PSI/3035/200/1}{Secure coding for medical
  sequence transmission over {ATM} network}, Medical Imaging 1997: PACS Design
  and Evaluation: Engineering and Clinical Issues 3035~(1) (1997) 200--209.
\newblock \href {http://dx.doi.org/10.1117/12.274572}
  {\path{doi:10.1117/12.274572}}.
\newline\urlprefix\url{http://link.aip.org/link/?PSI/3035/200/1}

\bibitem{Parrein2001}
B.~Parrein, P.~Verbert, N.~Normand, J.-P.~V. Gu\'edon,
  \href{http://link.aip.org/link/?PSI/4524/243/1}{Scalable multiple
  descriptions on packets networks via the n-dimensional {M}ojette transform},
  Quality of Service over Next-Generation Data Networks 4524~(1) (2001)
  243--252.
\newline\urlprefix\url{http://link.aip.org/link/?PSI/4524/243/1}

\bibitem{Verbert2002a}
P.~Verbert, J.-P.~V. Gu\'edon, B.~Parrein,
  \href{http://link.aip.org/link/?PSI/4862/315/1}{Distributed and compressed
  multimedia transmission using a discrete backprojection operator}, Internet
  Multimedia Management Systems III 4862~(1) (2002) 315--325.
\newblock \href {http://dx.doi.org/10.1117/12.473047}
  {\path{doi:10.1117/12.473047}}.
\newline\urlprefix\url{http://link.aip.org/link/?PSI/4862/315/1}

\bibitem{Guedon2001}
J.-P.~V. Gu\'edon, N.~Normand, P.~Verbert, B.~Parrein, F.~Autrusseau,
  \href{http://link.aip.org/link/?PSI/4519/226/1}{Load-balancing and scalable
  multimedia distribution using the {M}ojette transform}, Internet Multimedia
  Management Systems II 4519~(1) (2001) 226--234.
\newline\urlprefix\url{http://link.aip.org/link/?PSI/4519/226/1}

\bibitem{Autrusseau2002a}
F.~Autrusseau, J.~V. Gu\'edon,
  \href{http://link.aip.org/link/?PSI/4675/378/1}{Image watermarking for
  copyright protection and data hiding via the {M}ojette transform}, Security
  and Watermarking of Multimedia Contents IV 4675~(1) (2002) 378--386.
\newblock \href {http://dx.doi.org/10.1117/12.465296}
  {\path{doi:10.1117/12.465296}}.
\newline\urlprefix\url{http://link.aip.org/link/?PSI/4675/378/1}

\bibitem{Autrusseau2002}
F.~Autrusseau, J.~V. Gu\'edon,
  \href{http://dx.doi.org/10.1109/ICDSP.2002.1028193}{Image watermarking in the
  {F}ourier domain using the {M}ojette transform}, 2002 14th International
  Conference on Digital Signal Processing Proceedings. DSP 2002 (Cat.
  No.02TH8628) 2 (2002) 725--728.
\newline\urlprefix\url{http://dx.doi.org/10.1109/ICDSP.2002.1028193}

\bibitem{AutrusseauEtAl}
F.~Autrusseau, J.~Gu\'edon, Y.~Bizais,
  \href{http://dx.doi.org/10.1117/12.480296}{{M}ojette cryptomarking scheme for
  medical images}, Proc. of the SPIE - The International Society for Optical
  Engineering 5032 (2003) 958--965.
\newline\urlprefix\url{http://dx.doi.org/10.1117/12.480296}

\bibitem{Babel2005}
M.~Babel, B.~Parrein, O.~Deforges, N.~Normand, J.-P. Gu\'edon, J.~Ronsin,
  \href{http://link.aip.org/link/?PSI/5670/126/1}{Secured and progressive
  transmission of compressed images on the internet: application to
  telemedicine}, Internet Imaging VI 5670~(1) (2005) 126--136.
\newline\urlprefix\url{http://link.aip.org/link/?PSI/5670/126/1}

\bibitem{Servieres2003}
M.~Servi\`eres, J.-P. Gu\'edon, N.~Normand, A discrete tomography approach to
  {PET} reconstruction, in: Proc. 7th International Conf. on Fully 3D
  Reconstruction in Radiology and Nuclear Medicine, 2003.

\bibitem{Chandra2008}
S.~S. Chandra, I.~Svalbe, J.-P. Gu\'edon, An exact, non-iterative {M}ojette
  inversion technique utilising ghosts, in: Lecture Notes in Computer Science
  (LNCS), Vol. 4992, Springer Berlin / Heidelberg, 2008, pp. 401--412.

\bibitem{Fayad2008}
H.~Fayad, J.~P. Guedon, I.~Svalbe, Y.~Bizais, N.~Normand,
  \href{http://link.aip.org/link/?PSI/6913/69132S/1}{Applying {M}ojette
  discrete {R}adon transforms to classical tomographic data}, Medical Imaging
  2008: Physics of Medical Imaging 6913~(1) (2008) 69132S.
\newblock \href {http://dx.doi.org/10.1117/12.770478}
  {\path{doi:10.1117/12.770478}}.
\newline\urlprefix\url{http://link.aip.org/link/?PSI/6913/69132S/1}

\bibitem{Guedon2009}
J.-P. Gu\'edon, N.~Normand, A.~Kingston, B.~Parrein, M.~Servi\`eres,
  P.~\'Evenou, I.~Svalbe, F.~Autrusseau, T.~Hamon, Y.~Bizais, D.~Coeurjolly,
  F.~Boulos, E.~Grail, The {M}ojette {T}ransform: {T}heory and {A}pplications,
  ISTE-Wiley, 2009.

\bibitem{Jordan1870}
C.~Jordan, Trait\'e des substitutions et des \'equations alg\'ebriques,
  Gauthier-Villars, Paris, 1870.

\bibitem{Kung1979}
J.~P.~S. Kung,
  \href{http://www.sciencedirect.com/science/article/B6WHS-4D8DPY1-5N/2/2919cc%
9ecf355376df75bf47d0901506}{The {R}adon transforms of a {C}ombinatorial
  {G}eometry, {I}}, Journal of Combinatorial Theory, Series A 26~(2) (1979) 97
  -- 102.
\newblock \href {http://dx.doi.org/DOI: 10.1016/0097-3165(79)90059-1}
  {\path{doi:DOI: 10.1016/0097-3165(79)90059-1}}.
\newline\urlprefix\url{http://www.sciencedirect.com/science/article/B6WHS-4D8D%
PY1-5N/2/2919cc9ecf355376df75bf47d0901506}

\bibitem{Labunets1985}
V.~Labunets, Superfast multidimensional {F}ourier {R}adon transforms and
  multidimensional convolutions, Statistical Methods of Signal Processing IX
  (1985) 140--142.

\bibitem{Diaconis1985}
P.~Diaconis, R.~L. Graham, The {R}adon transform on $\mathbb{Z}^k_2$, Pacific
  Journal of Mathematics 118 (1985) 323--345.

\bibitem{Grigoryan1986}
A.~Grigoryan, New algorithms for calculating the discrete {F}ourier transforms,
  J. Vichislit. Matem. i Mat. Fiziki 25~(9) (1986) 1407--1412.

\bibitem{Bolker1987}
E.~D. Bolker, The {F}inite {R}adon {T}ransform, Contemporary Mathematics
  (American Mathematical Society) 63 (1987) 27--49.

\bibitem{Gertner1988}
I.~Gertner, A new efficient algorithm to compute the two-dimensional discrete
  {F}ourier transform, Acoustics, Speech and Signal Processing, IEEE
  Transactions on 36~(7) (1988) 1036--1050.
\newblock \href {http://dx.doi.org/10.1109/29.1627}
  {\path{doi:10.1109/29.1627}}.

\bibitem{Fill1989}
J.~A. Fill, \href{http://link.aip.org/link/?SJD/2/262/1}{The {R}adon
  {T}ransform on $\mathbb{Z}_n$}, SIAM Journal on Discrete Mathematics 2~(2)
  (1989) 262--283.
\newblock \href {http://dx.doi.org/10.1137/0402023}
  {\path{doi:10.1137/0402023}}.
\newline\urlprefix\url{http://link.aip.org/link/?SJD/2/262/1}

\bibitem{MatusFlusser}
F.~Mat\'u\v{s}, J.~Flusser, \href{http://dx.doi.org/10.1109/34.254058}{Image
  {R}epresentation via a {Fi}nite {R}adon {T}ransform}, Pattern Analysis and
  Machine Intelligence, IEEE Transactions on 15~(10) (1993) 996--1006.
\newline\urlprefix\url{http://dx.doi.org/10.1109/34.254058}

\bibitem{Cooley1965}
J.~W. Cooley, J.~W. Tukey, \href{http://dx.doi.org/10.2307/2003354}{An
  {A}lgorithm for the {M}achine {C}alculation of {C}omplex {F}ourier {S}eries},
  Mathematics of Computation 19~(90) (1965) 297--301.
\newblock \href {http://dx.doi.org/http://dx.doi.org/10.2307/2003354}
  {\path{doi:http://dx.doi.org/10.2307/2003354}}.
\newline\urlprefix\url{http://dx.doi.org/10.2307/2003354}

\bibitem{Rader1968}
C.~M. Rader, Discrete {F}ourier transforms when the number of data samples is
  prime, Proceedings of the IEEE 56~(6) (1968) 1107--1108.

\bibitem{Kingston2007}
A.~Kingston, I.~Svalbe,
  \href{http://www.sciencedirect.com/science/article/B6V09-4KF784N-1/2/8c70f35%
04e93d23e9b11e87981762893}{Generalised finite {R}adon transform for {N}x{N}
  images}, Image and Vision Computing 25~(10) (2007) 1620 -- 1630, {D}iscrete
  Geometry for Computer Imagery 2005.
\newblock \href {http://dx.doi.org/DOI: 10.1016/j.imavis.2006.03.002}
  {\path{doi:DOI: 10.1016/j.imavis.2006.03.002}}.
\newline\urlprefix\url{http://www.sciencedirect.com/science/article/B6V09-4KF7%
84N-1/2/8c70f3504e93d23e9b11e87981762893}

\bibitem{Chandra2010c}
S.~S. Chandra, Exact image representation via a {N}umber-{T}heoretic {R}adon
  {T}ransform, IET Computer Vision In Peer Review, submitted April.

\bibitem{Salzberg1999}
P.~M. Salzberg, R.~Figueroa, Tomography on the 3{D}-torus and crystals, in:
  G.~T. Herman, A.~Kuba (Eds.), Discrete Tomography: Foundations, Algortihms
  and Applications, Birkh\"{a}user, 1999, Ch.~19.

\bibitem{SvalbeVanDerSpek}
I.~Svalbe, D.~van~der Spek,
  \href{http://dx.doi.org/10.1016/S0024-3795(01)00487-6}{Reconstruction of
  tomographic images using analog projections and the digital {R}adon
  transform}, Linear Algebra and Its Applications 339 (2001) 125--145.
\newline\urlprefix\url{http://dx.doi.org/10.1016/S0024-3795(01)00487-6}

\bibitem{Svalbe2003}
I.~Svalbe, A.~Kingston,
  \href{http://www.sciencedirect.com/science/article/B75GV-4FWSP7W-7S/2/9f3640%
5de4b0cf6719fe4dac61e78f10}{Farey sequences and discrete {R}adon transform
  projection angles}, Electronic Notes in Discrete Mathematics 12 (2003) 154 --
  165, 9th International Workshop on Combinatorial Image Analysis.
\newblock \href {http://dx.doi.org/DOI: 10.1016/S1571-0653(04)00482-2}
  {\path{doi:DOI: 10.1016/S1571-0653(04)00482-2}}.
\newline\urlprefix\url{http://www.sciencedirect.com/science/article/B75GV-4FWS%
P7W-7S/2/9f36405de4b0cf6719fe4dac61e78f10}

\bibitem{Olds2000}
C.~Olds, A.~Lax, G.~Davidoff, The Geometry of Numbers, Vol.~41 of The Anneli
  Lax New Mathematical Library, The Mathematical Association of America, 2000.

\bibitem{Edholm1988}
P.~Edholm, G.~Herman, Linograms in image reconstruction from projections: Part
  1. back projecting using the {R}adon transform, Proc. Twenty-First Annual
  Hawaii International Conference on System Sciences 4 (1988) 48--50.

\bibitem{Kingston2003b}
A.~Kingston, I.~Svalbe, Mapping between digital and continuous projections via
  the discrete {R}adon transform in {F}ourier space, Proc. 7th International
  Conference on Digital Image Computing: Techniques and Applications, CSIRO
  Publishing (2003) 263--272.

\bibitem{Bube1997}
K.~P. Bube, R.~T. Langan, \href{http://link.aip.org/link/?GPY/62/1183/1}{Hybrid
  $\ell^1/ \ell^2$ minimization with applications to tomography}, Geophysics
  62~(4) (1997) 1183--1195.
\newblock \href {http://dx.doi.org/10.1190/1.1444219}
  {\path{doi:10.1190/1.1444219}}.
\newline\urlprefix\url{http://link.aip.org/link/?GPY/62/1183/1}

\bibitem{Candes2006}
E.~Cand\`{e}s, J.~Romberg, T.~Tao,
  \href{http://dx.doi.org/10.1109/TIT.2005.862083}{Robust uncertainty
  principles: exact signal reconstruction from highly incomplete frequency
  information}, Information Theory, IEEE Transactions on 52~(2) (2006)
  489--509.
\newblock \href {http://dx.doi.org/10.1109/TIT.2005.862083}
  {\path{doi:10.1109/TIT.2005.862083}}.
\newline\urlprefix\url{http://dx.doi.org/10.1109/TIT.2005.862083}

\bibitem{Press2006}
W.~H. Press, Discrete {R}adon transform has an exact, fast inverse and
  generalizes to operations other than sums along lines, Proceedings of the
  National Academy of Sciences (PNAS) 103~(51) (2006) 19249--19254.

\bibitem{AverbuchEtAl}
A.~Averbuch, R.~Coifman, D.~Donoho, M.~Israeli, J.~Walden, The pseudo-polar
  {FFT} and its applications, Technical Report YALEU/DCS/RR, Yale University
  1178.

\end{thebibliography}

\section*{Acronyms}
\begin{acronym}
\acrodef{1D}{one dimensional}
\acro{2D}{two dimensional}
\acro{RT}{Radon Transform}
\acro{CT}{Computerised Tomography}
\acro{FT}{Fourier Transform}
\acro{FST}{Fourier Slice Theorem}
\acro{DRT}{Discrete Radon Transform}
\acro{FRT}{Fast Radon Transform}
\acro{MT}{Mojette Transform}
\acro{FMT}{Fast Mojette Transform}
\acro{FFT}{Fast Fourier Transform}
\acro{DFT}{Discrete Fourier Transform}
\acro{NTT}{Number Theoretic Transform}
\acro{DPM}{Dirac Pixel Model}
\end{acronym}

\end{document}